\documentclass[%
twocolumn,reprint,
amsmath,amssymb,
aps,
pra,
]{revtex4-2}

\usepackage{graphicx}
\usepackage{dcolumn}
\usepackage{bm}
\usepackage{amsmath, amssymb}
\usepackage{makecell}
\usepackage{multirow}
\usepackage{float}

\def\pra{Phys. Rev. A}
\def\prb{Phys. Rev. B}
\def\prl{Phys. Rev. Lett.}
\def\ol{Opt. Lett.}
\def\opex{Opt. Express}
\def\nc{Nat. Commun.}
\def\np{Nat. Photon.}

\def\lpr{Laser Photonics Rev.}

\makeatletter

\newcommand{\Rmnum}[1]{\expandafter\@slowromancap\romannumeral #1@}
\makeatother                                                                   
\begin{document}                                                                
\preprint{APS/123-QED}

\title{Perturbation theory for resonant states near a bound state in the continuum}

\author{Nan Zhang}
\email{nzhang234-c@my.cityu.edu.hk}
\author{Ya Yan Lu}%

\affiliation{%
Department of Mathematics, City University of Hong Kong, Kowloon, Hong Kong, China
}%

\date{\today}

\begin{abstract} 
In this work, we develop a perturbation theory to analyze resonant states near a bound state in the continuum (BIC) in photonic crystal slabs.
The theory allows us to rigorously determine the asymptotic behavior of $Q$-factor and the far-field polarization.
We show that the resonant states close to a BIC can be nearly circularly 
polarized if the scattering matrix satisfies a certain condition. 
Moreover, our theory offers a novel perspective on super-BICs and provides a clear and precise condition to efficiently identify them in both symmetric and asymmetric structures without requiring a merging process.
For practical applications, we find a super-BIC in a square lattice of rods on a dielectric substrate.
Our theory addresses the non-Hermitian nature of the system, can be generalized to treat other structures that support BICs,
and has potential applications in resonance and chiral optics.
\end{abstract}

\maketitle

Bound states in the continuum (BICs) are characterized by their ability to maintain localization within a continuous spectrum of radiation states~\cite{Neumann1929PZ,Friedrich85PRA,Bonnet94MMAS,Evans94JFM,Marinica08PRL,Bulgakov08PRB,Plotnik11PRL,Hsu13Nature,Hsu16NRM,Kivshar2023PU}.
In a photonic crystal (PhC) slab, a BIC is surrounded by resonant states with $Q$-factor approaching infinity~\cite{Hsu13Nature,Hsu16NRM,Kivshar2023PU,Fan02PRB,Shipman03SIAMJAM,Lee12PRL}.
The high-$Q$ resonances and the related strong local fields~\cite{Mocella2015PRB} can be used to enhance light-matter interactions
for applications in solid-state lasing~\cite{Kodigala17Nature}, sensing~\cite{yanik2011PNAS}, and nonlinear optics~\cite{Yuan17PRA}, etc.
Merging multiple BICs to a special BIC, which is also referred to as a super-BIC by some authors~\cite{Yuan17PRA,Yuan18PRA,Zhen19Nature,Yuan20PRAPert,Kang21PRL,Yuri21NC,Kang22LSA,Luo23PRA,Bulgakov23PRB,Lee2023LPR,Liu2024OE,Le2024PRL,Zhang2024OL,Zhang2024OE},  
can improve $Q$-factor of nearby resonant states over a broad wavevector range
and suppress the radiation losses caused by fabrication imperfections~\cite{Zhen19Nature,Kang21PRL}.
In the momentum space, a BIC corresponds to a polarization singularity and is also a center of polarization vortex which 
has been observed experimentally~\cite{Doeleman2018NP,Zhang2018PRL}.
Such singularities can be used to manipulate light, giving rise to vortex beams generation and beam shifts~\cite{Wang2020NP,Wang2021NC}.
The far-field polarized states winding around the BIC induces a topological charge,
which can intuitively describe the evolution, generation, and annihilation of BICs~\cite{Zhen14PRL,Bulgakov17PRATopo,WZLiu19PRL,WLiu19PRL,Yoda20PRL,WLiu20LPR,Liu21Nano,Jiang2023PRL}. 

The far-field polarizations and $Q$-factor of resonant states near a BIC are crucial properties for understanding resonance-enhanced wave phenomena and exploring their applications. 
Existing studies on these two concepts are mostly numerical results, while some analytical results have been reported, they are tailored to specific models~\cite{Bulgakov23PRB,Zagaglia2023OL,Zhen19Nature,Zhao2019,Kazarinov1985} or particular cases~\cite{Yuan17PRA,Yuan18PRA}.
In this Letter, we develop a non-Hermitian perturbation theory for resonant states near a BIC in PhC slabs.
We show that if the scattering matrix is subject to a certain condition, 
the resonant states close to the BIC can exhibit near-circular polarization, 
which may have potential applications in chiral emission and chiroptical resonances~\cite{Zhang2022S,Liu2024OL,Chen2023Nature,Chen2022PRL,Zi2024PRL}.
Moreover, our theory offers a novel perspective on super-BICs and provides a clear and precise condition to find them efficiently in symmetric and asymmetric structures.
All existing examples of super-BICs have been calculated through a merging process~\cite{Zhen19Nature,Kang21PRL,Yuri21NC,Kang22LSA,Luo23PRA,Bulgakov23PRB,Lee2023LPR,Liu2024OE,Le2024PRL,Zhang2024OL,Zhang2024OE}.
Most of these examples require the maintenance of up-down mirror symmetry, with the exception of one particular study~\cite{Bai24}, which identified a super-BIC in a two-dimensional multilayer structure.
For practical applications, this merging process necessitates a substantial computational workload. 
In contrast, based on our theory, we identify a super-BIC in a relatively simple dielectric metasurface--a square lattice of rods on a substrate--without employing any merging process.

Our structure consists of a PhC slab sandwiched between two homogeneous media with dielectric constants $\varepsilon^\pm$,
where the symbols ``$\pm$'' indicate above and below the slab, respectively.
To simplify the presentation, we assume that the slab has a period $L$ in both $x$ and $y$,
and study resonant states near a non-degenerate BIC under the condition that only the zeroth diffraction order can propagate in the substrate and cladding.
We emphasize that our theory can accommodate more complicated scenarios, such as arbitrary periodic directions, 
multiple diffraction orders, and even degenerate cases~\cite{Bulgakov19,Bulgakov22PRB,Ochiai24,Zi2024PRL}.
The electric field of a resonant state can be written as ${\bm E}={\bm u}({\bm\rho},z)e^{i{\bm\alpha}\cdot{\bm\rho}}$, 
where ${\bm \rho}=(x,y)$, ${\bm\alpha}=(\alpha,\beta)$ is the Bloch wavevector,
${\bm u}$ is periodic in ${\bm \rho}$ and satisfies the wave equation
\begin{equation}\label{LetterMainEq}
	{\cal L}{\bm u}:=\left(\nabla+i{\bm\alpha}\right)\times\left(\nabla+i{\bm\alpha}\right) \times{\bm u}=\frac{\omega^2}{c^2}\varepsilon({\bm\rho},z){\bm u}. 
\end{equation} 
The far-field polarization vector of the resonant state in $sp$-plane is given by
${\bm d}^\pm=d_{s}^\pm{\hat{s}}+d_{p}^\pm{\hat p}^\pm$.

We use subscript ``$*$'' to denote the quantities of a BIC. 
In addition, we normalize the BIC by $\left<{\bm u}_*|\varepsilon|{\bm u}_*\right>=1$.
When the resonant state is close to the BIC, we can write ${\bm\alpha}$ as ${{\bm\alpha}_*}+\delta(\cos\theta,\sin\theta)/L$,
where $-\pi\leq \theta<\pi$ and $\delta$ is a small positive parameter.
If $\delta$ is sufficiently small, the frequency $\omega$, the periodic state ${\bm u}$ and the operator ${\cal L}$ can be expanded by
\begin{align}
	\label{Letterperfre}
	&\omega=\omega_*+\delta \omega_1(\theta)+\delta^2 \omega_2(\theta)+\cdots\\
	\label{Letterperu}
	&{\bm u}={\bm u}_*+\delta {\bm u}_1(\theta)+\delta^2 {\bm u}_2(\theta)+\cdots\\
	\label{Letterperope}
	&{\cal L}={\cal L}_*+\delta{\cal L}_1(\theta)+\delta^2{\cal L}_2(\theta),
\end{align}
where ${\cal L}_1$ and ${\cal L}_2$ are operators depending on the angle $\theta$ and can be written down explicitly.

Substituting Eqs.~(\ref{Letterperfre})-(\ref{Letterperope}) into Eq.~(\ref{LetterMainEq})
and collecting terms with the same order, we obtain a sequence of recursive equations.
For each $\theta$, 
the far-field pattern of the 1st-order state correction ${\bm u}_1$ is exactly a plane wave with the polarization vector
${\bm d}_1^\pm=d_{1s}^\pm {\hat{s}_*}+d_{1p}^\pm {\hat{p}}_*^\pm$. 
We show that the second-order radiation loss of resonant states near the BIC depends on the outgoing power of the plane wave, i.e.,
\begin{equation}\label{Letter2ndRL}
	2\omega_*\mbox{Im}(\omega_2)L=-c^2\left(\gamma_*^+|{\bm d}_1^+|^2+\gamma_*^-|{\bm d}_1^-|^2\right),
\end{equation}
where $\gamma_*^\pm=\sqrt{\omega_*^2\varepsilon^\pm/c^2-|{\bm\alpha}_*|^2}$.

The polarization vector ${\bm d}_1^\pm$ can be determined by involving the scattering states (diffraction solutions)  corresponding to the BIC.
When the slab is illuminated by $s$ and $p$ polarized incident plane waves in the upper and lower regions, 
we obtain four linearly independent scattering states ${\bm u}_{s*}^{\pm}$ and ${\bm u}_{p*}^{\pm}$. 
If the scattering states and the BIC are orthogonal, i.e., $\left<{\bm u}_{l*}^{\pm}|\varepsilon|{\bm u}_*\right>=0$, $l\in\{s,p\}$,
${\bm d}_1^\pm$ can be calculated from
\begin{equation}\label{Letterd1sd1p}
	\begin{aligned}
		\left[
		\begin{array}{c}
			d_{1s}^+\\
			d_{1p}^+\\
			d_{1s}^-\\
			d_{1p}^-			
		\end{array}
		\right]={\bm \Gamma}^{-1/2}{\bf S}{\bf U}\left[
		\begin{array}{c}
			\cos\theta\\
			\sin\theta
		\end{array}
		\right]
	\end{aligned},
\end{equation}
where ${\bm \Gamma}=\mbox{diag}(\gamma_*^+,\gamma_*^+,\gamma_*^-,\gamma_*^-)L$, $\bf S$ is the scattering matrix, and $\bf U$ is a $4\times 2$ matrix given by
\begin{equation}
	{\bf U}=\left[
	\begin{array}{c}
		{\bf U}^+\\
		{\bf U}^-
	\end{array}
	\right],\;{\bf U}^\pm=\left[
	\begin{array}{cc}
		U_{sx}^\pm&U_{sy}^\pm\\
		U_{px}^\pm&U_{py}^\pm
	\end{array}
	\right].
\end{equation}
Here $U_{l\sigma}^\pm=L^2\left<{\bm u}_{l*}^\pm|{\cal L}_{1\sigma}|{\bm u}_*\right>$ for $l\in\{s,p\}$ and $\sigma\in\{x,y\}$.
The operators ${\cal L}_{1\sigma}$ are defined from the decomposition ${\cal L}_1(\theta)={\cal L}_{1x}\cos\theta + {\cal L}_{1y}\sin\theta$.
Notice that matrices $\bf S$ and $\bf U$ depend solely on the BIC.
The detailed derivation of Eqs.~(\ref{Letter2ndRL}) and (\ref{Letterd1sd1p}) is given in the Supplemental
Material~\cite{NanSM}\nocite{Sakoda2012OE,LJYuan2017,Zhang2024arXiv}.

From Eqs.~(\ref{Letterperfre})-(\ref{Letterd1sd1p}), 
it is evident that if ${\bm d}_1^\pm\neq 0$, we have $Q\sim 1/\delta^2$ and ${\bm d}^\pm\sim\delta{\bm d}_1^\pm$ as $\delta\rightarrow 0$.
If ${\bm d}_1^\pm=0$, the asymptotic behavior of the $Q$-factor and the polarization depends on higher order state corrections.
In that case, we have $Q\sim 1/\delta^4$ at least and the BIC is a super-BIC.
Consequently, our theory offers a new perspective on super-BICs,
namely, the ultrahigh-$Q$ resonances are caused by a nonradiating 1st-order state correction
and are independent of any merging process.
Furthermore, the above analysis does not rely on any in-plane symmetry of the structure and the theory is general.

Many theoretical studies and applications favor structures with an in-plane inversion symmetry, 
since then many BICs can exist robustly~\cite{Zhen14PRL}.
In the following, we consider a PhC slab with this symmetry, 
and scale the BIC such that ${\cal C}_2\overline{\bm u}_*={\bm u}_*$,
where ${\cal C}_2$ is a $180^{\circ}$ rotation operator around the $z$ axis~\cite{Zhen14PRL,Sakoda2005Book}.
We show that the matrix ${\bf V}={\bf S}^{1/2}{\bf U}$ is real,
and if $\bf V$ is full rank, then ${\bm d}_1^\pm\neq 0$ for all $\theta$.
We refer to such a BIC as a generic BIC.
A super-BIC has a vanishing ${\bm d}_1^\pm$ at all or some angles, corresponding to a rank-deficient $\bf V$~\cite{NanSM}.
More precisely, super-BICs can be categorized into two classes:

{\em class 1}: ${\bf V}=0$. We have ${\bm d}_1^\pm=0$ and $Q\sim 1/\delta^{4}$ at least for all $\theta$.
The ultrahigh-$Q$ resonances occur in all directions.

{\em class 2}: $\mbox{rank}({\bf V})=1$. We only have two angles $\theta_s$ such that $[\cos\theta_s,\sin\theta_s]^{\sf T}$ belong to the null space of $\bf V$. 
For $\theta\neq \theta_s$, we have 
${\bm d}_1^\pm\neq 0$ and $Q\sim 1/\delta^2$.
For $\theta= \theta_s$, we have 
${\bm d}_1^\pm=0$ and $Q\sim 1/\delta^{4}$ at least. 
The ultrahigh-$Q$ resonances occur only in these two directions,
and the asymptotic growth order of $Q$-factor exhibits directionality.

\begin{figure}[htbp]
	\centering
	\includegraphics[scale=0.125]{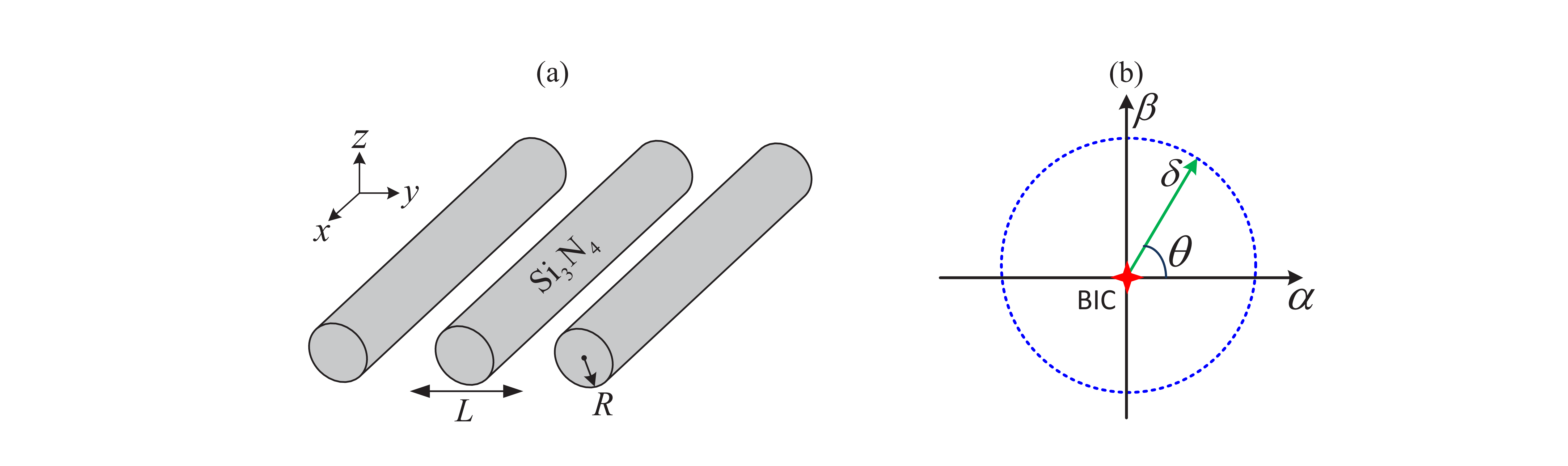}\\
	\vspace{0.5cm}
	\includegraphics[scale=0.19]{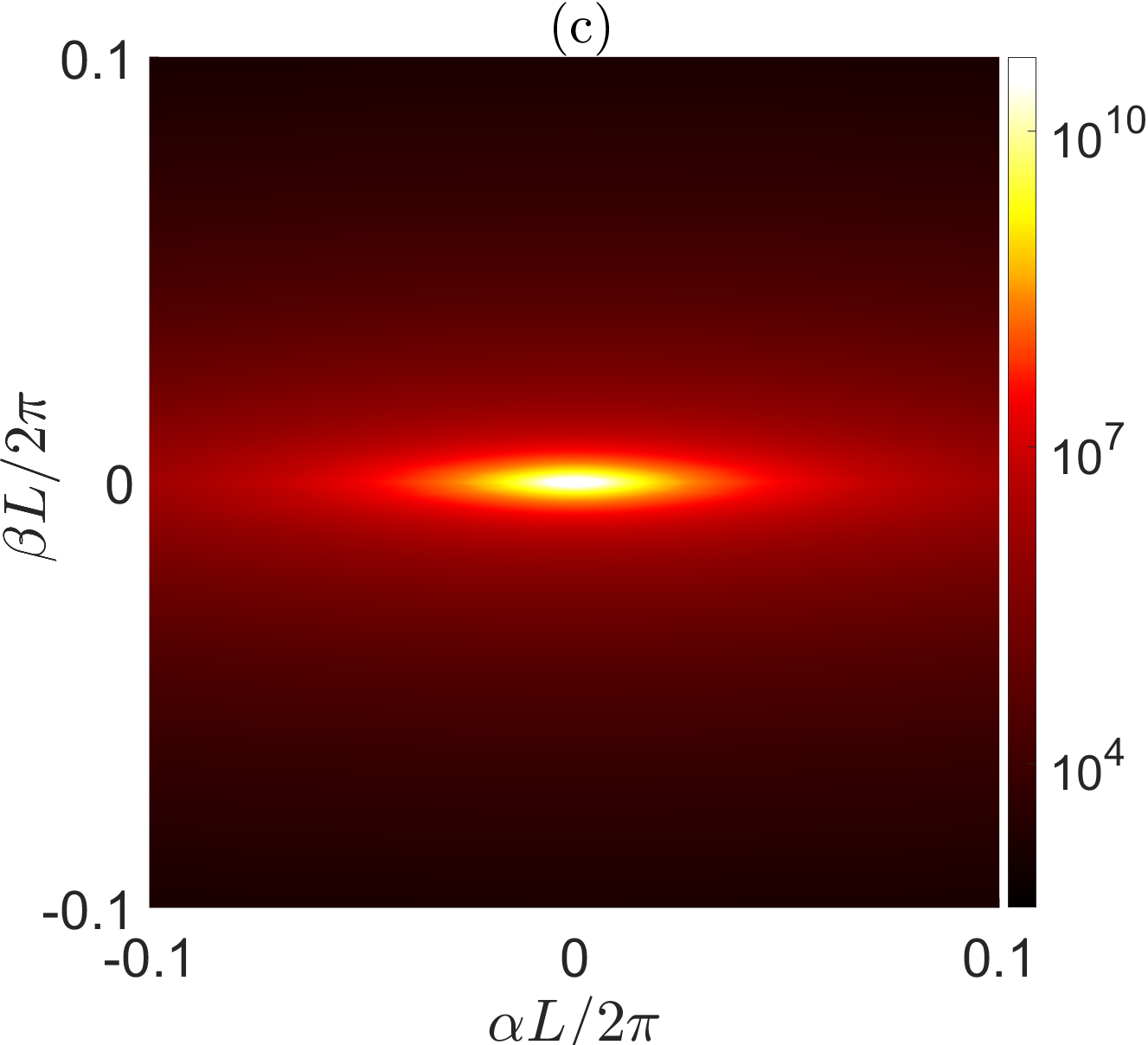}
	\includegraphics[scale=0.19]{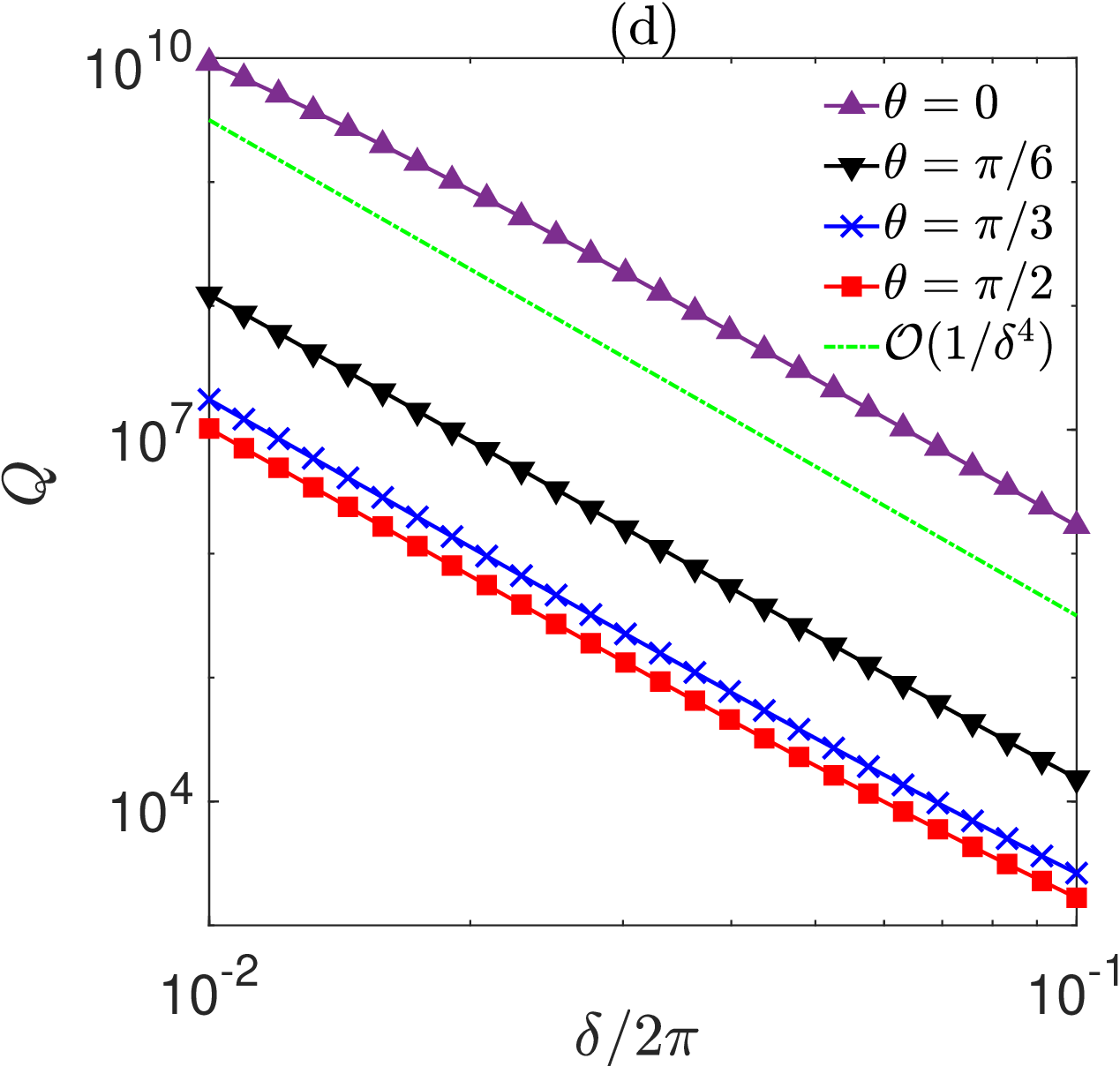}\\
	\vspace{0.5cm}
	\includegraphics[scale=0.19]{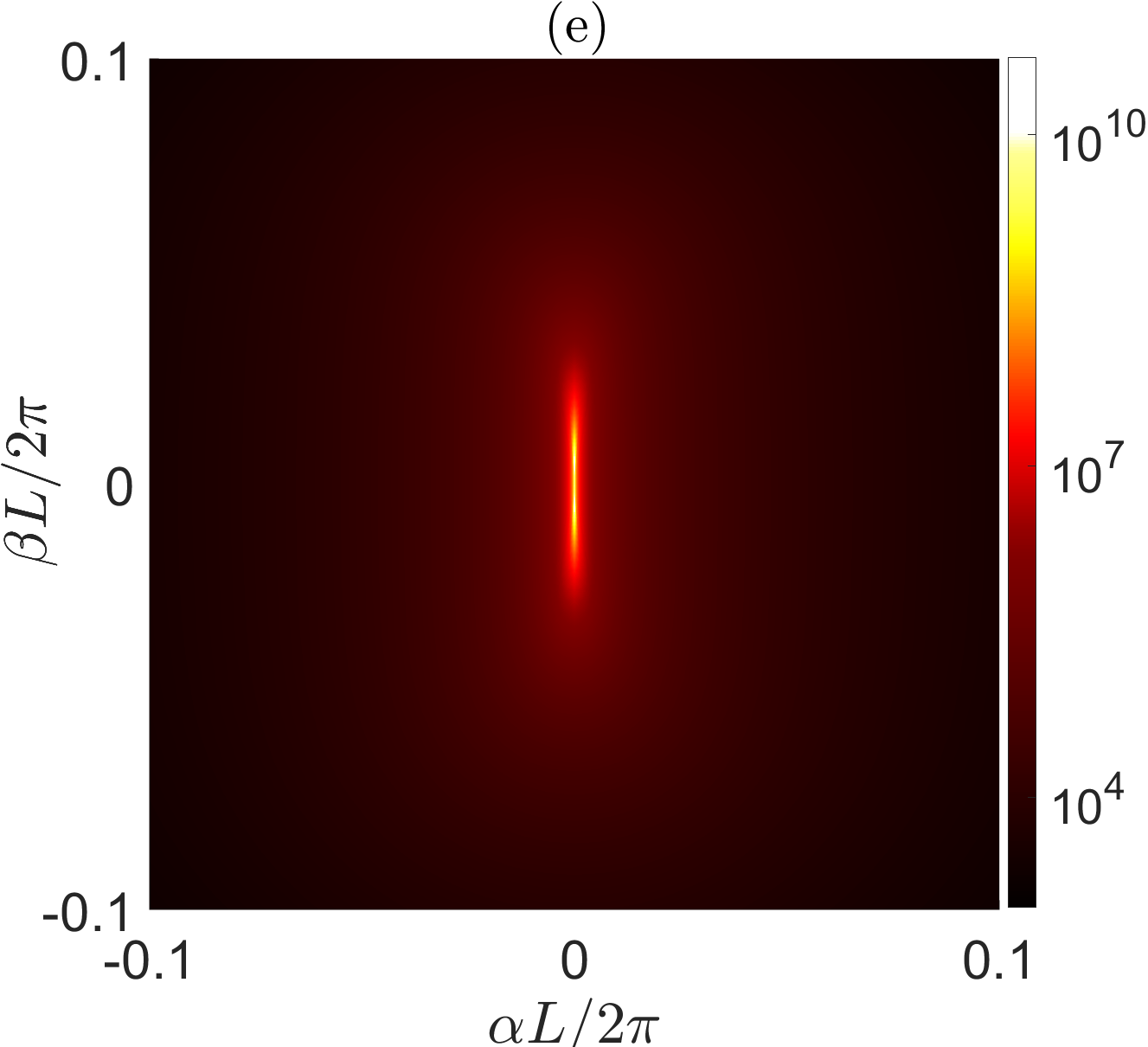}
	\includegraphics[scale=0.19]{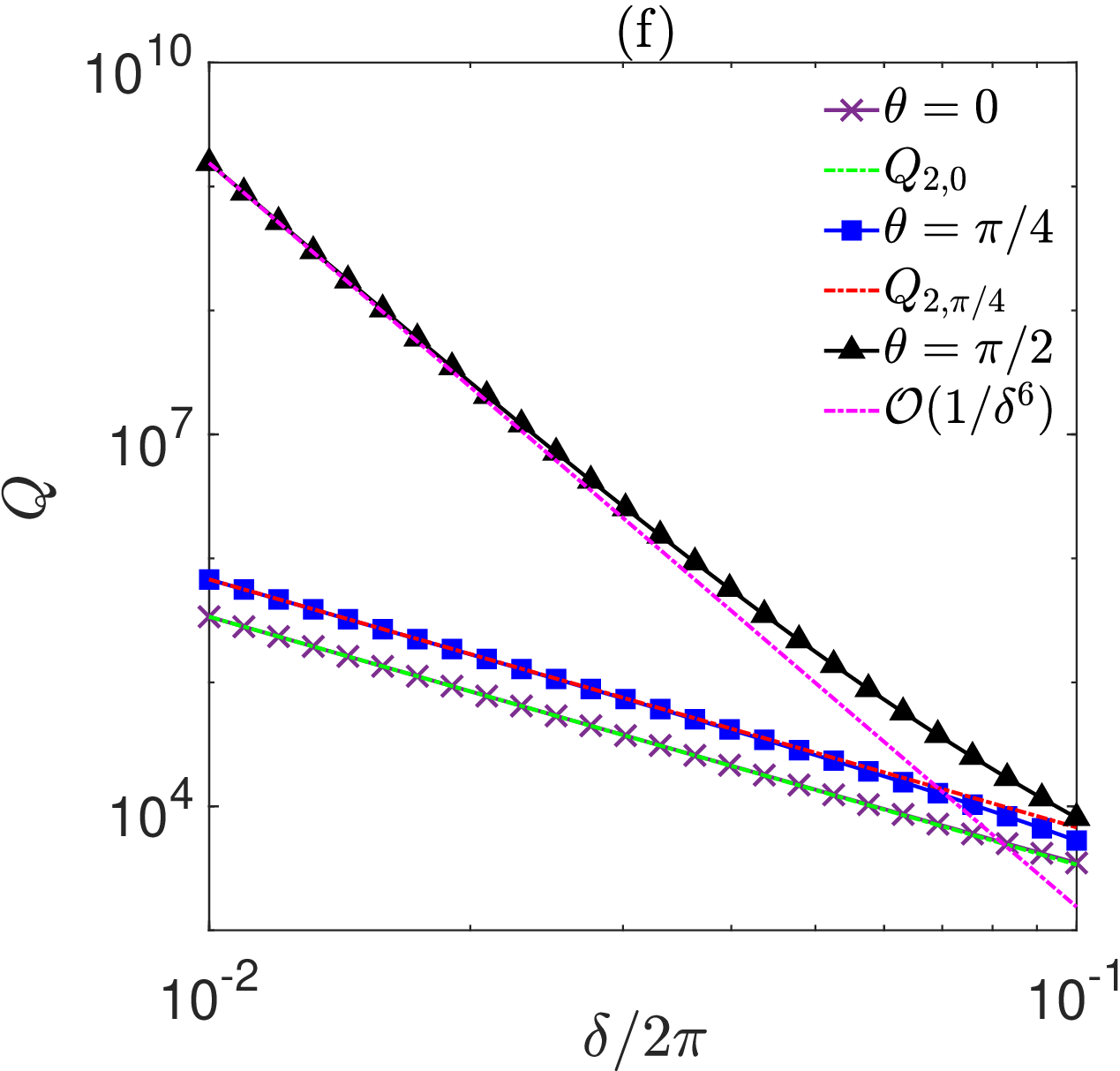}
	\vspace{0.2cm}
	\caption{The $Q$-factor of resonant states near two at-$\Gamma$ super-BICs in a periodic array of circular cylinders.  			
		(a). Schematic of the structure ($\varepsilon_{{\rm Si}_3{\rm N}_4}=4$). 
		(b). The momentum space or $\alpha$-$\beta$ plane.	
		(c) and (d). A super-BIC ($\omega_*L/2\pi c=0.677$ and ${\cal C}_2{\bm u}_*=-{\bm u}_*$)  belongs to {\em class 1} with ultrahigh-$Q$ resonances occurring at all angles.
		(e) and (f). A super-BIC ($\omega_*L/2\pi c=0.562$ and ${\cal C}_2{\bm u}_*={\bm u}_*$) belongs to {\em class 2} with ultrahigh-$Q$ resonances existing only at two special angles $\theta_s=\pm \pi/2$.
		The values of $Q$ calculated numerically are nearly identical with the predicted results $Q_{2,\theta}$.
	}\label{LetterPACylinderQFactor}
\end{figure}

We provide several numerical examples of at-$\Gamma$ BICs to verify our theory.
Some results of off-$\Gamma$ BICs can be founded in the Supplemental
Material~\cite{NanSM}.
For an at-$\Gamma$ BIC, the related scattering states can be scaled such that ${\cal C}_2{\bm u}_{l*}^\pm=-{\bm u}_{l*}^\pm$.
If the BIC is not symmetry-protected (${\cal C}_2{\bm u}_*=-{\bm u}_*$),
according to the anticommutativity of operators ${\cal L}_{1\sigma}$ and ${\cal C}_2$, i.e., 
${\cal L}_{1\sigma}{\cal C}_2=-{\cal C}_2{\cal L}_{1\sigma}$,
we have $U_{l\sigma}^\pm=0$ and obtain a zero ${\bf V}$.
In this case, we typically have $Q\sim 1/\delta^4$ for all $\theta$ and this BIC is a super-BIC belonging to {\em class 1}. 
As an example, we consider the TE mode in a periodic array of silicon nitride circular cylinders shown in Fig.~\ref{LetterPACylinderQFactor}(a).
The cylinders are infinitely long in $x$ and are surrounded by air. 
At the radius $R=0.398L$, we obtain a BIC with ${\cal C}_2{\bm u}_*=-{\bm u}_*$.
Figure~\ref{LetterPACylinderQFactor}(c) and (d) show that we indeed have $Q\sim1/\delta^4$ for all $\theta$.
In addition, we consider a BIC belonging to the irreducible {\em B} representation in a PhC slab with ${C}_{6v}$ symmetry~\cite{Zhen14PRL,Sakoda2005Book}.
Such a BIC also satisfies ${\cal C}_2{\bm u}_*=-{\bm u}_*$ and we then have $Q\sim 1/\delta^4$ in all directions at least.

A symmetry-protected BIC typically is generic and has a full rank matrix $\bf V$.
It becomes a super-BIC if the matrix $\bf V$ is rank-deficient.
Considering the structure shown in Fig.~\ref{LetterPACylinderQFactor}(a), the matrix $\bf V$ of a BIC holds the form
\begin{equation}\label{LetterFormV}
	{\bf V}=\left[
	\begin{array}{cc}
		0&V_{12}\\
		V_{21}&0\\
		0&V_{32}\\
		V_{41}&0
	\end{array}
	\right].
\end{equation}
In addition, we have $V_{12}=V_{32}$ and $V_{21}=V_{41}$ due to up-down mirror symmetry.
At $R=0.439L$, there exists a super-BIC with $V_{12}=0$ and $V_{21}\neq 0$.
From Eq.~(\ref{Letterd1sd1p}), we can obtain ${\bm d}_1^\pm=0$ for $\theta_s=\pm\pi/2$.
Moreover, since the BIC is symmetry-protected, 
we can further prove that $Q\sim 1/\delta^6$ at least~\cite{NanSM}. 
For other $\theta$, we can calculate the vectors ${\bm d}_1^\pm$ from Eq.~(\ref{Letterd1sd1p}) and approximate $Q$-factor by $Q_{2,\theta}=-0.5\omega_*/\mbox{Im}[\delta^2\omega_2(\theta)]$.
Figure~\ref{LetterPACylinderQFactor}(e) and (f) demonstrate the validity of our theory. 
\begin{figure}[htbp]
	\centering
	\includegraphics[scale=0.21]{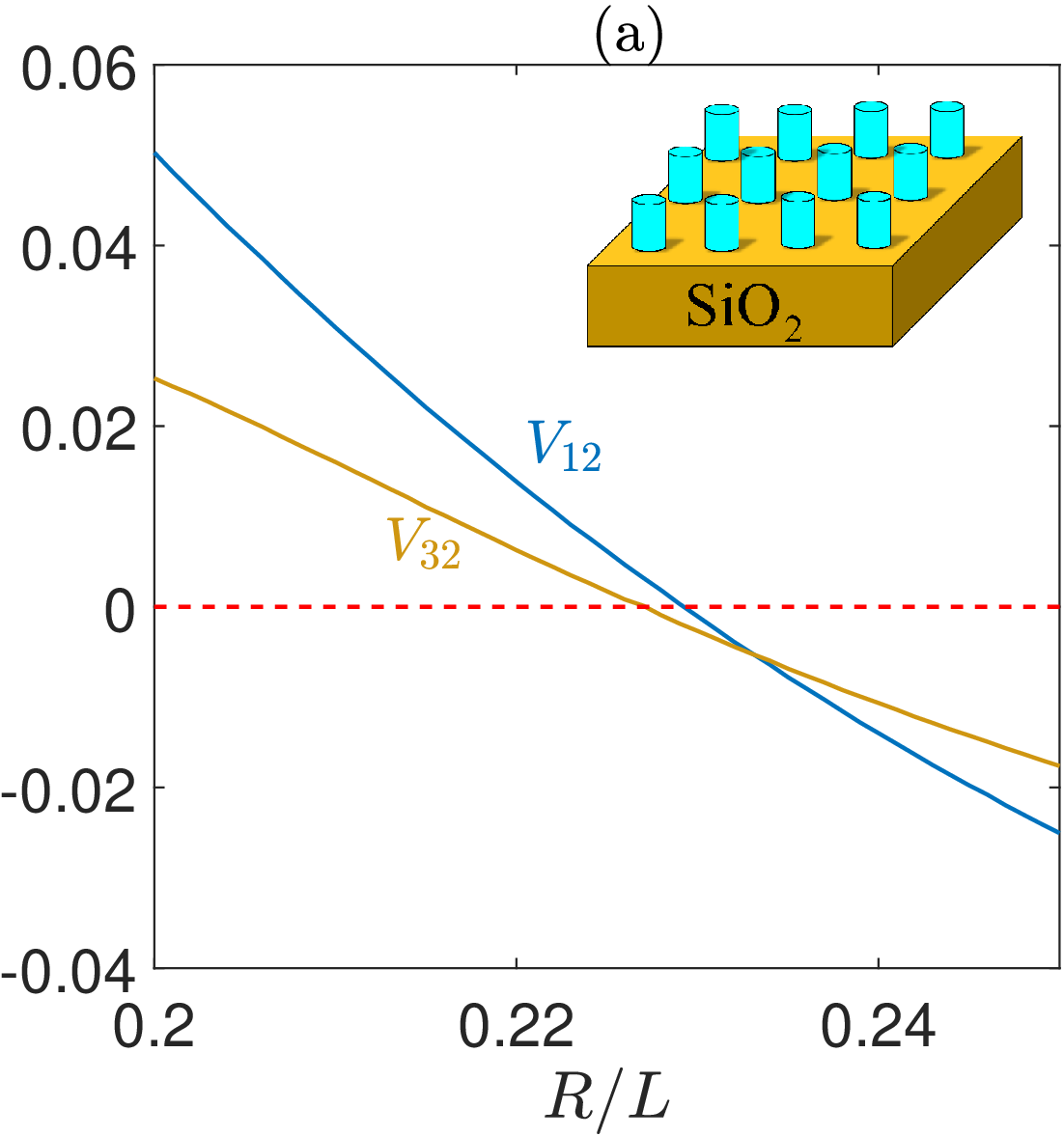}\hspace{0.55cm}
	\includegraphics[scale=0.21]{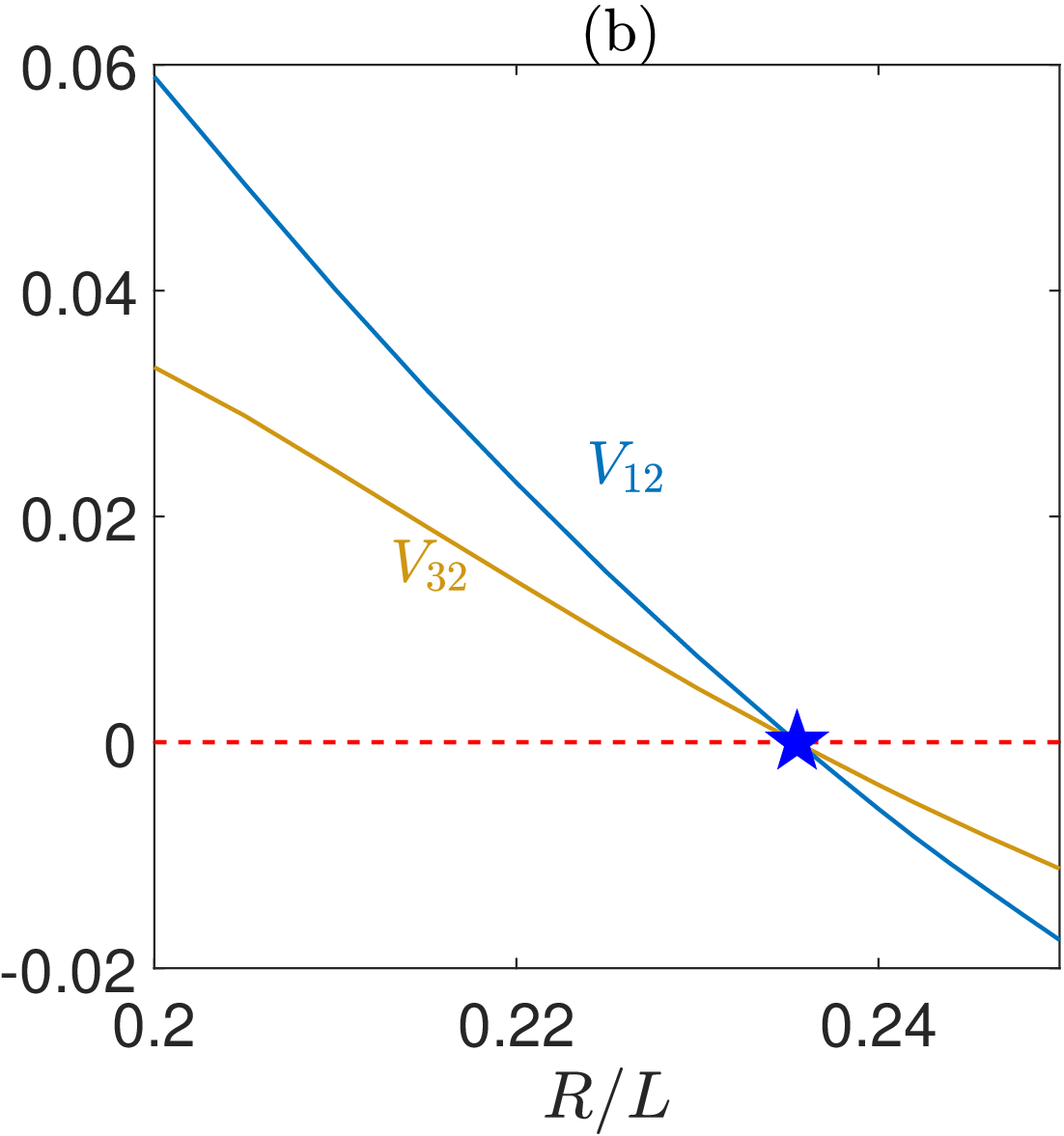}\\
	\vspace{0.2cm}
	\includegraphics[scale=0.21]{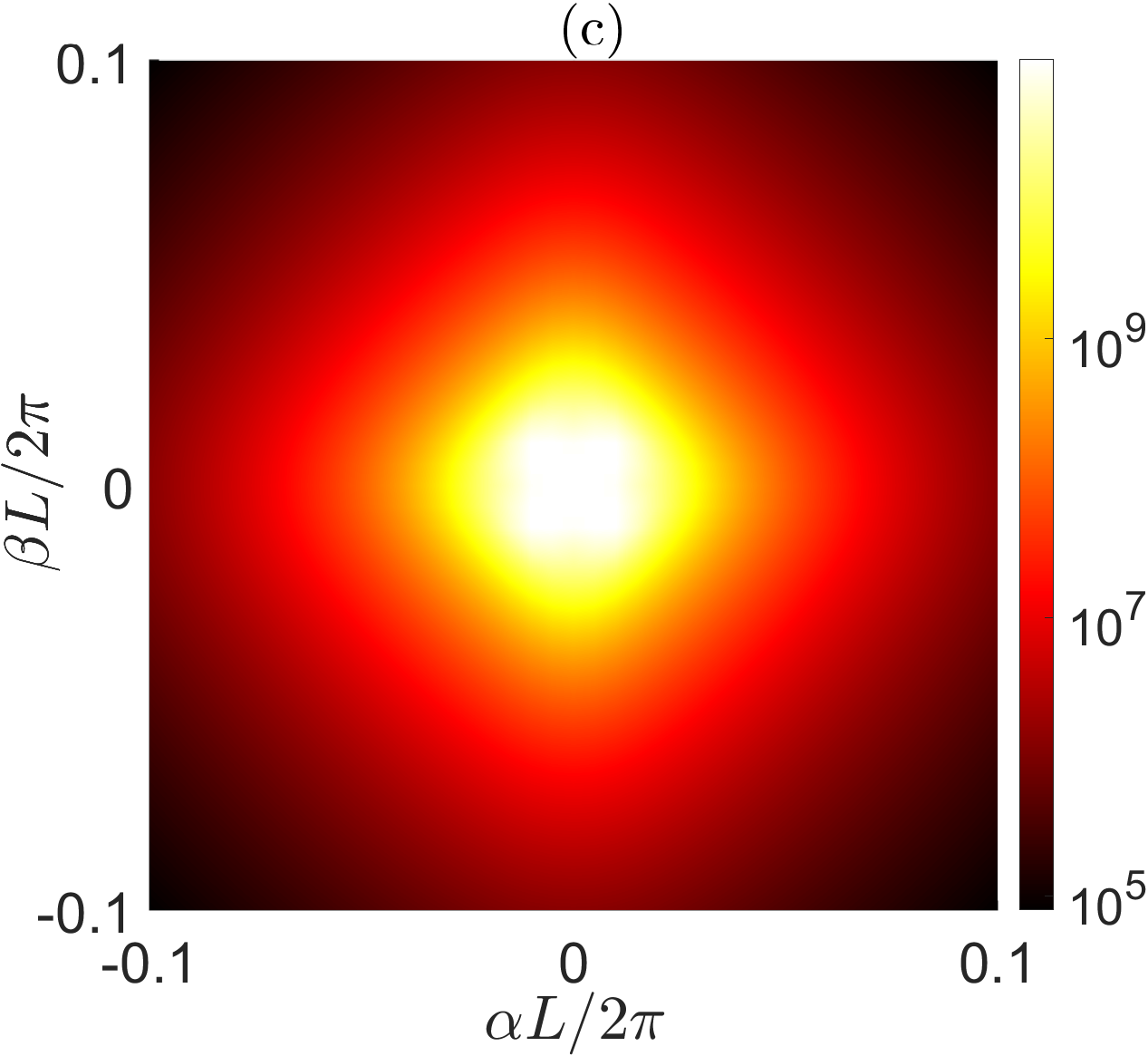}
	\includegraphics[scale=0.21]{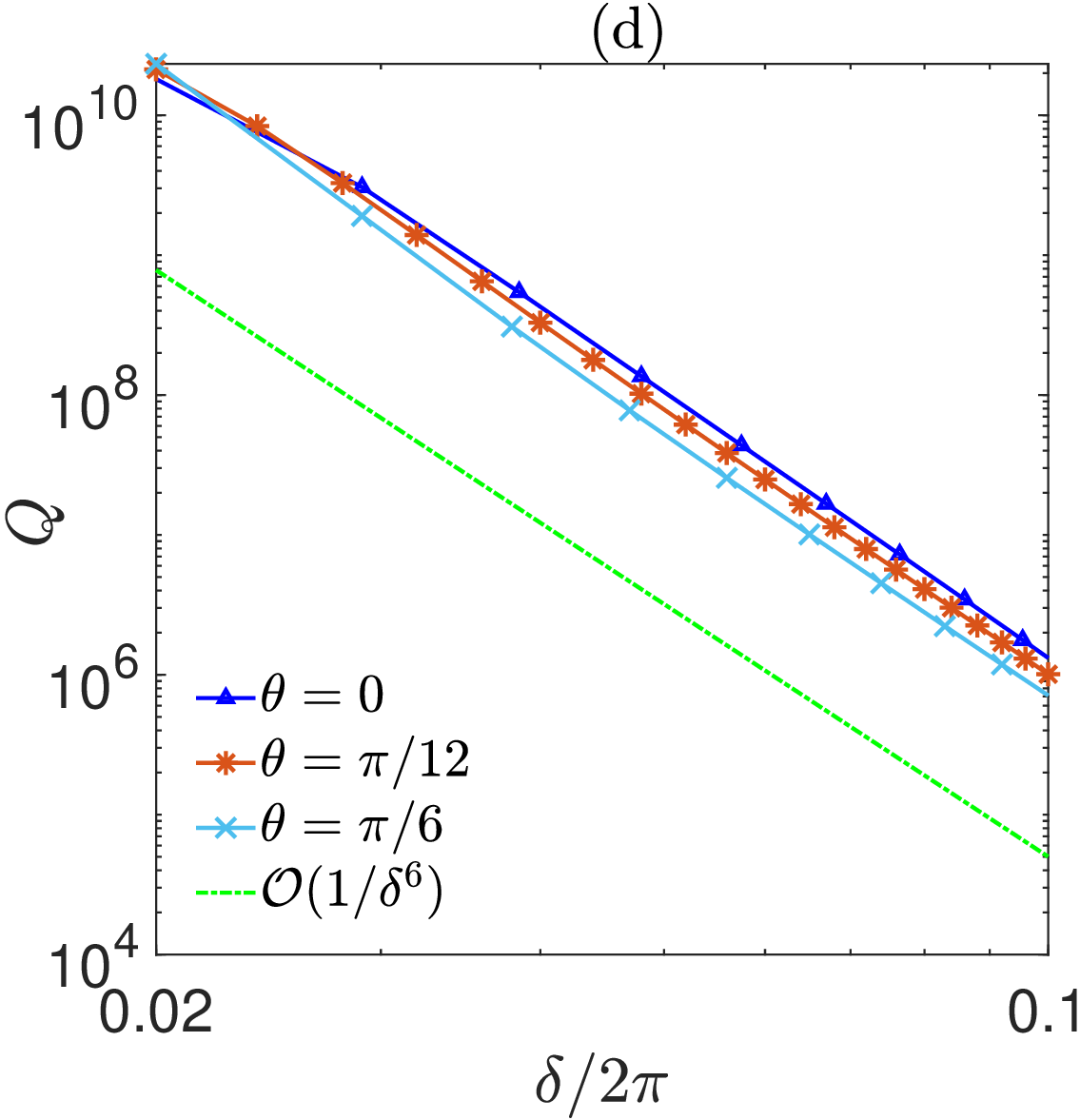}
	\vspace{0.2cm}
	\caption{A super-BIC in a square lattice of silicon rods on a silica substrate ($\varepsilon_{{\rm SiO}_2}=2.1025$). (a) and (b). The quantities $V_{12}$ and $V_{21}$ as functions of radius $R$ for heights $h=0.8L$ and $h=0.8205L$, respectively. The blue pentagram corresponds to the super-BIC. (c). $Q$-factor (calculated numerically) of resonant states near the super BIC ($\omega_*L/2\pi c=0.512$) at $R=0.236L$ and $h=0.8205L$. (d). Asymptotic behavior of $Q$-factor in three directions.}\label{AsymmetricSuperBIC}
\end{figure}

For BICs in structures with $C_{4v}$ symmetry, 
the related scattering states belong to the irreducible {\em E} representation.
We consider BICs belonging to the irreducible $A_2$ representation
and show that the matrix $\bf V$ holds the form in Eq.~(\ref{LetterFormV}) with $V_{12}=-V_{21}$ and $V_{32}=-V_{41}$.
A super-BIC corresponds to $V_{12}=V_{32}=0$.
In this case, we can further prove that $Q\sim 1/\delta^6$ for all $\theta$ at least~\cite{NanSM}. 
To validate our theory, we consider a square lattice of silicon rods in air.
Due to up-down mirror symmetry, we only need to tune radius $R$ to obtain a zero of $V_{12}=V_{32}$.
As shown in Fig.~\ref{SMSymSuperBIC}~\cite{NanSM}, for a fixed height $h=0.8L$, we have $V_{12}=0$ and obtain a super-BIC at $R=0.221L$. 

If the structure breaks up-down mirror symmetry, 
we have $V_{12}\neq V_{32}$ and usually need to tune two structural parameters to find common zeros of $V_{12}$ and $V_{32}$.
As an example, we consider a square lattice of silicon rods on a silica substrate.
The quantities $V_{12}$ and $V_{32}$ as functions of radius $R$ for two values of height, $h=0.8L$ and $0.8205L$, are shown in Fig.~\ref{AsymmetricSuperBIC}(a) and (b).
Compared with the previous example,
for the same height $h=0.8L$, 
$V_{12}$ and $V_{32}$ are different and their zeros also do not coincide.
When the height $h$ is increased to $0.8205L$, 
zeros of $V_{12}$ and $V_{32}$ coincide at $R=0.236L$ and we obtain a super-BIC (marked by a blue pentagram).
As shown in Fig.~\ref{AsymmetricSuperBIC}(c)-(d), we have $Q\sim 1/\delta^6$ in all directions.
The above results indicate that based on our theory, 
super-BICs can be efficiently identified even in the absence of up-down mirror symmetry.
However,
finding such a super-BIC using a merging approach is remarkably challenging, 
as it requires the tuning of a parameter to find off-$\Gamma$ BICs, 
and further tuning to merge them with an at-$\Gamma$ BIC. 
The existing examples of super-BICs in structures breaking up-down mirror symmetry are limited to two-dimensional configurations and involve specific structural characteristics~\cite{Bai24}. Moreover, as we have previously demonstrated, these super-BICs exhibit ultrahigh-$Q$ resonances only at two specific angles. 
In contrast, our structure is relatively simple, and ultrahigh-$Q$ resonances occur at all angles.

The preceding analysis has examined the asymptotic behavior of $Q$-factor.
In the following, we study the far-field polarization of resonant states near generic BICs.
Our theory applies to super-BICs as well.
For simplicity, we consider a structure exhibiting 
up-down mirror symmetry and replace half of the structure with a perfect electric or magnetic conductor wall, 
consistent with the up-down mirror symmetry of ${\bm u}_*$.
The far-field polarization vectors of resonant states $\bm u$ and 1st-order state corrections ${\bm u}_1$
are denoted by ${\bm d}$ and ${\bm d}_1$, respectively.
The vector ${\bm d}_1$ can be calculated from Eq.~(\ref{Letterd1sd1p}) in which the scattering matrix $\bf S$ is given by
\begin{equation}
	{\bf S}=\left[
	\begin{array}{cc}
		C_{ss}&C_{ps}\\
		C_{sp}&C_{pp}
	\end{array}
	\right].
\end{equation}
Here $C_{l\sigma}$ denotes the reflection coefficient of the $l$-polarized 
incident wave reflected into a $\sigma$-polarized wave, where $l,\sigma\in\{s,p\}$.
The polarization angles for ${\bm d}$ and ${\bm d}_1$ are defined by 
$\phi=\arg(\mathbb{S}_{1}+i\mathbb{S}_{2})/2$ and $\phi_1=\arg(\mathbb{S}_{1,1}+i\mathbb{S}_{1,2})/2$, 
and the ellipticities are defined by $\chi=\mathbb{S}_3/\mathbb{S}_0$ and $\chi_1=\mathbb{S}_{1,3}/\mathbb{S}_{1,0}$, 
where $\mathbb{S}_m$ and $\mathbb{S}_{1,m}$, $m=0,1,2,3$, are Stokes parameters for ${\bm d}$ and ${\bm d}_1$, respectively.
It is clear that $\phi\sim \phi_1$ and $\chi\sim\chi_1$ as $\delta\rightarrow 0$.

Whether there are circularly polarized states arbitrarily close to a BIC is an important theoretical question~\cite{Zhen14PRL}.
In the following, we show that there exist some BICs where ${\bm d}_1$ is circularly polarized at some angles $\theta_c$.
In these directions, the resonant states close to the BIC are nearly circularly polarized.
As shown in the Supplemental
Material~\cite{NanSM}, 
circularly polarized ${\bm d}_1$ can occur if and only if the scattering coefficients satisfy $\arg(C_{ss}\overline{C}_{pp})=\pm\pi$ or $C_{ss}=C_{pp}=0$ (polarization conversion). The angles $\theta_c$ can be solved from the relation $d_{1s}=\pm id_{1p}$ and Eq.~(\ref{Letterd1sd1p}).

\begin{figure}[htbp]
	\centering
	\includegraphics[scale=0.25]{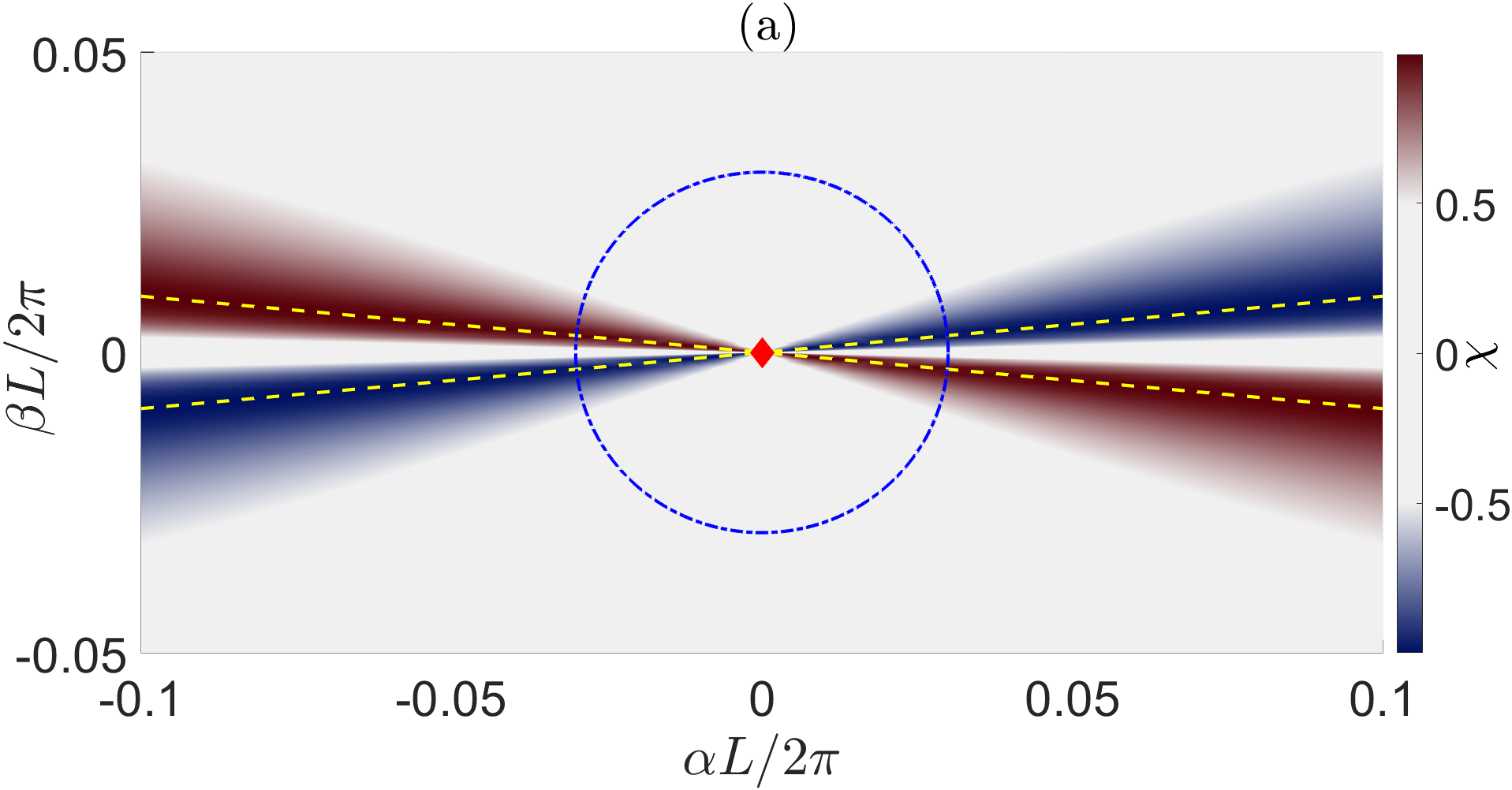}\\
	\vspace{0.2cm}
	\includegraphics[scale=0.225]{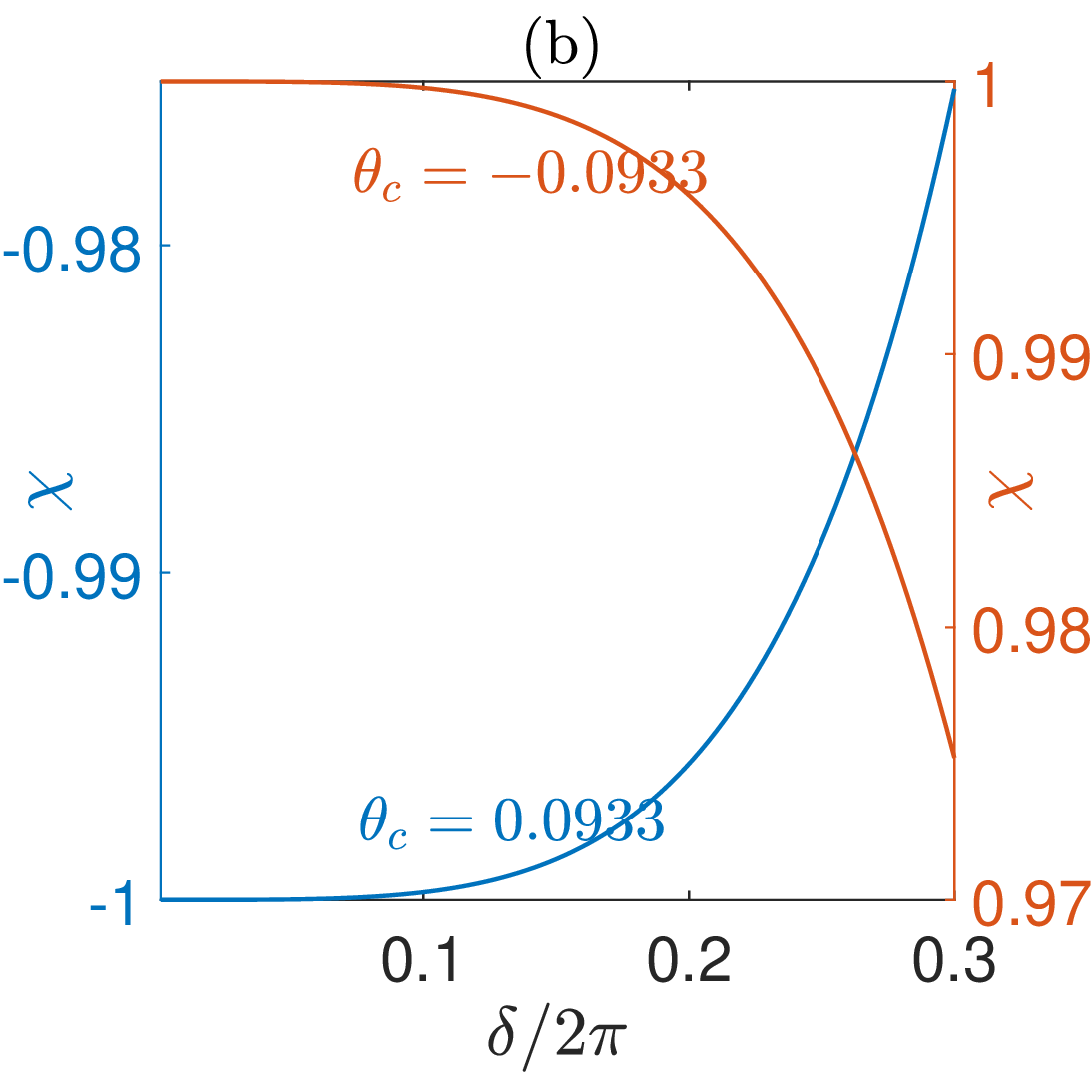}\,
	\includegraphics[scale=0.225]{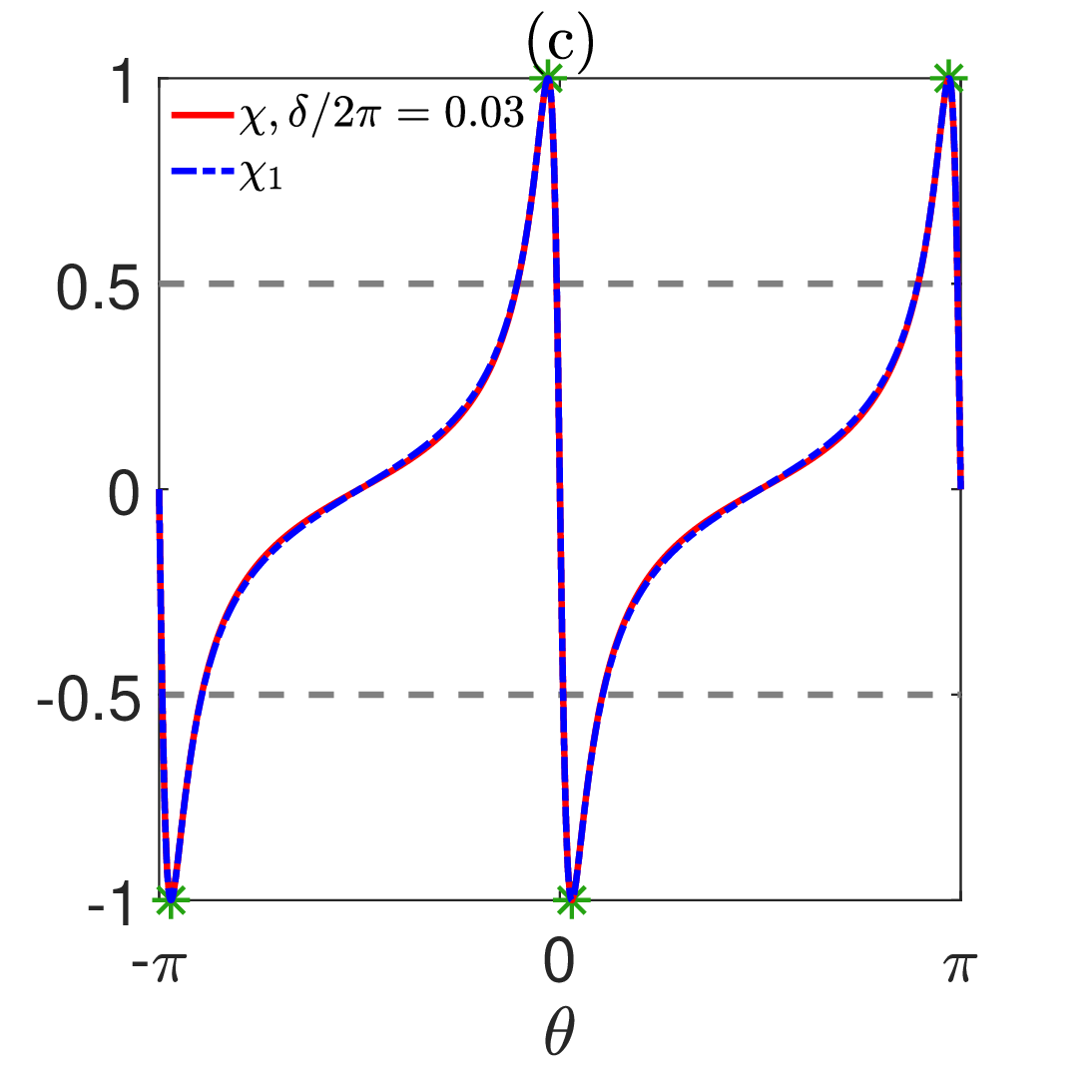}	
	\vspace{0.2cm}
	\caption{(a). The ellipticity $\chi$ in momentum space. The red diamond represents the BIC ($\omega_*L/2\pi c=0.851$). 
		The yellow dash lines represent angles $\theta_c=\pm 0.0933$ and $\theta_c=\pm 0.0933\mp\pi$.
		Notice that resonant states with $|\chi|>0.5$ can exist in four cone regions (dark colored regions).
		The blue dot-dashed circle denotes resonant states with ${\bm\alpha}=\delta(\cos\theta,\sin\theta)/L$,
		where $\delta/2\pi = 0.03$.
		(b). The ellipticity $\chi$ versus $\delta$ at $\theta_c=\pm 0.0933$, respectively. As $\delta\rightarrow 0$,
		we have $|\chi|\rightarrow 1$.	
		(c). The ellipticity $\chi$ of resonant states at the blue dot-dashed circle shown in panel (a).
		It is clear that the values of $\chi$ calculated numerically are nearly identical with the predicted results $\chi_1$.
		The green stars denote the angles $\theta_c$ at which $\chi_1=\pm 1$.
	}\label{LetterS3S0CPS}
\end{figure}
To validate our theory, we consider the TE mode in the structure shown in Fig.~\ref{LetterPACylinderQFactor}(a). 
At $R=0.2L$, we have a BIC with $\arg(C_{ss}\overline{C}_{pp})=\pi$.
A straightforward calculation gives
four angles $\theta_c$ at which ${\bm d}_1$ is circularly polarized,
i.e., $\chi_1=\pm 1$.
Figure~\ref{LetterS3S0CPS}(a) shows the ellipticity $\chi$ in momentum space,
revealing the presence of resonant states with $|\chi|>0.5$ in four conical regions (dark-colored regions) induced by these special angles $\theta_c$ (yellow lines).
Along these angles, as illustrated in Fig.~\ref{LetterS3S0CPS}(b), we indeed have $|\chi|\rightarrow 1$ as $\delta\rightarrow 0$.
Moreover, as shown in Fig.~\ref{LetterS3S0CPS}(a) and (c),
the ellipticity $\chi$ of resonant states at the blue dot-dashed circle closely match 
$\chi_1$.
The cone angle of these four conical regions can be estimated from $\chi_1$ directly.
These results imply that our theory can approximate the polarizations efficiently, predict the existence and determine the corresponding angles of nearly circular polarizations.

If ${\bm d}_1$ is not circularly polarized for all directions, 
we show that the matrices ${\bf S}$ and ${\bf V}$ can indicate the topological nature of the BIC directly.
The Stokes parameters $\mathbb{S}_{1,1}$, $\mathbb{S}_{1,2}$ can be written in a compact form
\begin{equation}\label{LetterStokes}
	\begin{aligned}
		\left[
		\begin{array}{c}
			\mathbb{S}_{1,1}\\
			\mathbb{S}_{1,2}
		\end{array}
		\right]={\bf F}\left[
		\begin{array}{c}
			\cos 2\theta\\
			\sin 2\theta
		\end{array}
		\right]+{\bf g},
	\end{aligned}
\end{equation}
where $\bf F$ is a real matrix and $\bf g$ is a real vector obtained from the definition of Stokes parameters and Eq.~(\ref{Letterd1sd1p}) directly.
The matrix $\bf F$ is non-singular if and only if ${\bm d}_1$ is not circularly polarized for all $\theta$~\cite{NanSM}. 
In this case, the trajectory $(\mathbb{S}_{1,1},\mathbb{S}_{1,2})$ forms an ellipse with two revolutions around the origin. 
The revolution direction depends on the matrices ${\bf S}$ and ${\bf V}$.
Therefore, we must have $\phi_1(\pi)-\phi_1(-\pi)=\pm 2\pi$ and obtain topological charge $\pm 1$ directly from $\phi_1$.

Finally, we discuss the polarizations near an at-$\Gamma$ BIC in a structure with $C_{4v}$ symmetry. 
In this case, the phase difference between $d_{1s}$ and $d_{1p}$ is 0 or $\pi$, i.e., ${\bm d}_1$ is linearly polarized. 
Therefore, resonant states near an at-$\Gamma$ BIC in a structure with $C_{4v}$ symmetry are nearly linearly polarized.
A general condition applicable to other BICs regarding the existence of near-linear polarizations is outlined in the Supplemental
Material~\cite{NanSM}.

In summary, we developed a perturbation theory to analyze resonant states near a BIC in PhC slabs, 
and determined the asymptotic behavior of $Q$-factor and the polarization rigorously.
Our theory introduces a novel perspective on super-BICs and provides a clear and precise condition 
for efficiently identifying them in both symmetric and asymmetric structures, 
without the need to follow a merging process. 
As an illustrative example, we found a super-BIC in a square lattice of silicon rods on a silica substrate.
Additionally, we have demonstrated that under a certain condition, the resonant states near the BIC can exhibit near-circular polarization. 
The findings presented in this paper may hold direct and potential applications in the fields of resonance and chiral optics.
Although the current study has focused on resonant states near a BIC in PhC slabs, 
our theory can be extended to other systems where BICs exist~\cite{Gao19ACS,Bulgakov17PRA,Zhang23OE,webster07PTL,zou15LPR,Yuan21OE,Gomis17NP}.

The authors acknowledge support from the Research Grants Council of Hong
Kong Special Administrative Region, China (Grant No. CityU 11317622). 

\bibliography{apssamp}

\onecolumngrid

\renewcommand{\thefigure}{S\arabic{figure}}
\renewcommand{\thesection}{S\arabic{section}}
\renewcommand{\theequation}{S\arabic{equation}}

\indent

\begin{center}\large
	\textbf{Supplementary Material}
\end{center}

\section{Derivations of equations (5) and (6)}

The wave equation for resonant states can be written as 
\begin{equation}\label{MainEq}
	{\cal L}{\bm u}=k^2\varepsilon({\bm\rho},z){\bm u}, 
\end{equation}
where $k=w/c$ is the free space wavenumber, $\omega$ is the angular frequency, and $c$ is the speed of light in vacuum. 
The wavenumber of resonant states near a BIC can be expanded as
\begin{equation}\label{perk}
	k=k_*+\delta k_1(\theta)+\delta^2 k_2(\theta)+\cdots.
\end{equation}
Substituting Eqs.~(3)-(4) and (\ref{perk}) into Eq.~(\ref{MainEq}) and collecting the ${\cal O}(\delta)$ terms, 
we obtain the 1st-order perturbation equation
\begin{equation}\label{LEq1}
	{\cal M}_*{\bm u}_1=\left[2k_*k_1\varepsilon-{\cal L}_1(\theta)\right]{\bm u}_*,
\end{equation}
where ${\cal M}_*={\cal L}_*-k_*^2\varepsilon$ is the wave operator.
Regarding Eq.~(\ref{LEq1}) as a radiation problem and according to the asymptotic behavior of Green's function, we have
\[
{\bm u}_1\sim{\bm d}_1^\pm e^{\pm i\gamma_{*}^\pm z},\; z\rightarrow\pm \infty.
\]
With Green's identities, we have 
\begin{equation}\label{k1realcond}
	\left<{{\bm u}}_*|{\cal M}_*|{{\bm u}}_1\right>:=\frac{1}{L^3}\int_\Omega\overline{{\bm u}}_*\cdot{\cal M}_*{{\bm u}}_1\,{\rm d}x{\rm d}y{\rm d}z=0,
\end{equation}
where $\overline{\bm u}_*$ denotes the complex conjugate of ${\bm u}_*$ and $\Omega=(-L/2,L/2)\times(-L/2,L/2)\times\mathbb{R}$ is a unit cell.
For convenience, we restate Green's identities here:
\[
\iint_V\left({\bm w} \cdot \nabla \times \nabla \times {\bm v}-{\bm v} \cdot \nabla  \times \nabla \times {\bm w}\right)\;{\rm d}V =\int_{S}\left({\bm v} \times \nabla \times {\bm w}-{\bm w} \times \nabla \times {\bm v}\right) \cdot {\bm n}\; {\rm d} S,
\]
and
\[
\iint_V\left({\bm w} \cdot \nabla \times  {\bm v}-{\bm v} \cdot \nabla  \times{\bm w}\right)\;{\rm d}V =\int_{S}({\bm n}\times {\bf v}) \cdot {\bf u}\;{\rm d} S,
\]
where $V$ is a domain with the boundary $S$, ${\bm n}$ is the outward unit normal vector of $S$, ${\bm w}$ and ${\bm v}$ are vector functions.

Equation~(\ref{k1realcond})
leads to $2k_*k_1=\left<{{\bm u}}_*|{\cal L}_1|{{\bm u}}_*\right>$,
which is simply the Hellmann–Feynman theorem.
By Green's identities, we can show that $\left<{{\bm u}}_*|{\cal L}_1|{{\bm u}}_*\right>$ is real 
and obtain a real $k_1$.
Substituting Eqs.~(3)-(4) and (\ref{perk}) into Gauss's law $(\nabla+i{\bm \alpha})\cdot {\bm u}=0$ and collecting the ${\cal O}(\delta)$ terms, 
we obtain ${\bm k}_*^\pm\cdot{\bm d}_1^\pm=0$, where ${\bm k}_*^\pm=(\alpha_*,\beta_*,\pm\gamma_*^\pm)$.
Thus the vectors ${\bm d}_1^\pm$ can be decomposed as
$
{\bm d}_1^\pm=d_{1s}^\pm {\hat{s}}_*+d_{1p}^\pm {\hat{p}}_*^\pm,
$
where, 
\[
\hat{s}_*=\left\{\begin{aligned}
	&\frac{{{\hat{z}}\times {\bm k}_*^\pm}}{{|{\hat{z}}\times {\bm k}_*^\pm|}},&&{\hat{z}}\times {\bm k}_*^\pm\neq 0\\
	&{\hat x},&&{\hat{z}}\times{\bm k}_*^\pm=0
\end{aligned}
\right.,\quad
{\hat{p}}_*^\pm=\pm{\bm k}_*^{\pm}\times{\hat{s}_*}/k_*.
\]
Collecting the ${\cal O}(\delta^2)$ terms in Eq.~(\ref{MainEq}), we can obtain
\begin{equation}\label{LEq2}
	{\cal M}_*{\bm u}_2=\left(2k_*k_1\varepsilon-{\cal L}_1\right){\bm u}_1+[\left(k_1^2+2k_*k_2\right)\varepsilon-{\cal L}_2]{\bm u}_*.
\end{equation}
Similarly, we can show that $\left<{{\bm u}}_*|{\cal M}_*|{{\bm u}}_2\right>=0$, which gives rise to
\[
\mbox{Im}(2k_*k_2)=-\mbox{Im}\left<{{\bm u}}_*\left|{2k_*k_1}\varepsilon-{\cal L}_1\right|{\bm u}_1\right>=\mbox{Im}\left<{{\bm u}}_1\left|{\cal M}_*\right|{\bm u}_1\right>.
\]
Since 
$
\mbox{Im}\left<{{\bm u}}_1\left|{\cal M}_*\right|{\bm u}_1\right>
=\left(\left<{{\bm u}}_1\left|{\cal L}_*\right|{\bm u}_1\right>-
\overline{\left<{{\bm u}}_1\left|{\cal L}_*\right|{\bm u}_1\right>}\right)/2i,
$
according to the asymptotic behavior of ${\bm u}_1$, we obtain Eq.~(5) in the Letter.

To calculate ${\bm d}_1^\pm$, we introduce four linearly independent diffraction solutions.
We denote several plane waves by
\[
{\bm s}^\pm=\eta^\pm\hat{s}e^{\pm i\gamma_*^\pm z},\;{\bm p}^\pm_s=\eta^\pm{\hat p}^\pm e^{\pm i\gamma_*^\pm z},\;{\bm p}^\pm_i=-{\cal C}_2{\bm p}_s^\pm,
\]
where $\eta^\pm=0.5/\sqrt{\gamma_*^\pm L}$ are used to normalize the power.
When the structure is illuminated by incident plane waves $\overline{\bm s}^\pm$ and $\overline{\bm p}^\pm_i$ in the upper and lower regions, respectively,
we obtain four linearly independent diffraction solutions $\tilde{\bm u}^{\pm}_{s*}$ and $\tilde{\bm u}^{\pm}_{p*}$ satisfying the asymptotic behavior
\[
\tilde{\bm u}^{+}_{s*}\sim\left\{
\begin{aligned}
	&\overline{\bm s}^+ + R_{ss}^+{\bm s}^+ + R_{ps}^+{\bm p}^+_s,\;z\rightarrow+\infty\\
	&T_{ss}^+{\bm s}^- + T_{ps}^+{\bm p}^-_s,\;z\rightarrow-\infty
\end{aligned}
\right.,\quad
\tilde{\bm u}^{+}_{p*}\sim\left\{
\begin{aligned}
	&\overline{\bm p}^+_i + R_{sp}^+{\bm s}^+ + R_{pp}^+{\bm p}^+_s,\;z\rightarrow+\infty\\
	&T_{sp}^+{\bm s}^- + T_{pp}^+{\bm p}^-_s,\;z\rightarrow-\infty
\end{aligned}
\right.
\]
\[
\tilde{\bm u}^{-}_{s*}\sim\left\{
\begin{aligned}
	&T_{ss}^-{\bm s}^+ + T_{ps}^-{\bm p}^+_s,\;z\rightarrow+\infty\\
	&\overline{\bm s}^-+ R_{ss}^-{\bm s}^- + R_{ps}^-{\bm p}^-_s,\;z\rightarrow-\infty
\end{aligned}
\right.,\quad
\tilde{\bm u}^{-}_{p*}\sim\left\{
\begin{aligned}
	&T_{sp}^-{\bm s}^+ + T_{pp}^-{\bm p}^+_s,\;z\rightarrow+\infty\\
	&\overline{\bm p}^-_i+R_{sp}^-{\bm s}^- + R_{pp}^-{\bm p}^-_s,\;z\rightarrow-\infty
\end{aligned}
\right..
\]
Since ${\cal M}_*\tilde{\bm u}^\pm_{l*}=0$, $l\in\{s,p\}$, we have
\[
\left<\tilde{\bm u}^\pm_{l*}|{\cal M}_*|{\bm u}_1\right>=\left<\tilde{\bm u}^\pm_{l*}|{\cal M}_*|{\bm u}_1\right>-\overline{\left<{\bm u}_1|{\cal M}_*|\tilde{\bm u}^\pm_{l*}\right>}=
\left<\tilde{\bm u}^\pm_{l*}|{\cal L}_*|{\bm u}_1\right>-\overline{\left<{\bm u}_1|{\cal L}_*|\tilde{\bm u}^\pm_{l*}\right>}.
\]
A straightforward calculation shows that
\[
\begin{aligned}
	&iL^2\left<\tilde{\bm u}^+_{l*}|{\cal M}_*|{\bm u}_1\right>=\sqrt{\gamma_*^+ L}(\overline{R}_{sl}^+d_{1s}^++\overline{R}_{pl}^+d_{1p}^+)+\sqrt{\gamma_*^- L}(\overline{T}_{sl}^+d_{1s}^-+\overline{T}_{pl}^+d_{1p}^-)\\
	&iL^2\left<\tilde{\bm u}^-_{l*}|{\cal M}_*|{\bm u}_1\right>=\sqrt{\gamma_*^+ L}(\overline{T}_{sl}^-d_{1s}^++\overline{T}_{pl}^-d_{1p}^+)+\sqrt{\gamma_*^- L}(\overline{R}_{sl}^-d_{1s}^-+\overline{R}_{pl}^-d_{1p}^-),
\end{aligned}
\]
where $l\in\{s,p\}$.
We denote $i\tilde{\bm u}_{l*}^\pm$ by ${\bm u}_{l*}^\pm$ and rewrite the above equations in a compact form
\[
\left[
\begin{array}{c}
	d_{1s}^+\\
	d_{1p}^+\\
	d_{1s}^-\\
	d_{1p}^-			
\end{array}
\right]=-L^2{\bf \Gamma}^{-1/2}{\bf S}\left[
\begin{array}{c}
	\left<{\bm u}_{s*}^+|{\cal M}_*|{\bm u}_1\right>\\
	\left<{\bm u}_{p*}^+|{\cal M}_*|{\bm u}_1\right>\\
	\left<{\bm u}_{s*}^-|{\cal M}_*|{\bm u}_1\right>\\
	\left<{\bm u}_{p*}^-|{\cal M}_*|{\bm u}_1\right>
\end{array}
\right],
\]
where ${\bm \Gamma}=\mbox{diag}(\gamma_*^+,\gamma_*^+,\gamma_*^-,\gamma_*^-)L$ and $\bf S$ is the scattering matrix given by
\[
{\bf S}=\left[
\begin{array}{cccc}
	{R}_{ss}^+&{R}_{sp}^+&{T}_{ss}^-&{T}_{sp}^-\\
	{R}_{ps}^+&{R}_{pp}^+&{T}_{ps}^-&{T}_{pp}^-\\
	{T}_{ss}^+&{T}_{sp}^+&{R}_{ss}^-&{R}_{sp}^-\\
	{T}_{ps}^+&{T}_{pp}^+&{R}_{ps}^-&{R}_{pp}^-
\end{array}
\right].
\]
If we shift the diffraction solutions such that they are orthogonal with the BIC, i.e.,
$
\left<{\bm u}^{\pm}_{l*}|\varepsilon |{\bm u}_*\right> =0,
$                                   
according to Eq.~(\ref{LEq1}), we obtain Eq.~(6) in the Letter, i.e.
\[
\left[
\begin{array}{c}
	d_{1s}^+\\
	d_{1p}^+\\
	d_{1s}^-\\
	d_{1p}^-			
\end{array}
\right]=L^2{\bm \Gamma}^{-1/2}{\bf S}\left[
\begin{array}{c}
	\left<{\bm u}_{s*}^+|{\cal L}_1|{\bm u}_*\right>\\
	\left<{\bm u}_{p*}^+|{\cal L}_1|{\bm u}_*\right>\\
	\left<{\bm u}_{s*}^-|{\cal L}_1|{\bm u}_*\right>\\
	\left<{\bm u}_{p*}^-|{\cal L}_1|{\bm u}_*\right>
\end{array}
\right].
\]

For a super-BIC with vanishing ${\bm d}_1^\pm$, by using the same procedure, we obtain 
\[
{\bm u}_2\sim{\bm d}_2^\pm e^{\pm i\gamma_{*}^\pm z},\; z\rightarrow\pm \infty.
\]
We also have ${\bm k}^\pm_*\cdot{\bm d}_2^\pm=0$ and the vectors ${\bm d}_2^\pm$ can be calculated from
\begin{equation}\label{d2sd2p}
	\begin{aligned}
		\left[
		\begin{array}{c}
			d_{2s}^+\\
			d_{2p}^+\\
			d_{2s}^-\\
			d_{2p}^-			
		\end{array}
		\right]=-L^2{\bm \Gamma}^{-1/2}{\bf S}\left[
		\begin{array}{c}
			\left<{\bm u}_{s*}^+|{\cal M}_*|{\bm u}_2\right>\\
			\left<{\bm u}_{p*}^+|{\cal M}_*|{\bm u}_2\right>\\
			\left<{\bm u}_{s*}^-|{\cal M}_*|{\bm u}_2\right>\\
			\left<{\bm u}_{p*}^-|{\cal M}_*|{\bm u}_2\right>
		\end{array}
		\right].
	\end{aligned}
\end{equation}
If ${\bm d}_2^\pm\neq 0$, we have 
$2k_*\mbox{Im}(k_4)L=-\left(\gamma_*^+|{\bm d}_2^+|^2+\gamma_*^-|{\bm d}_2^-|^2\right)$.
As $\delta\rightarrow 0$, we have $Q\sim 1/\delta^4$ and ${\bm d}\sim\delta^2{\bm d}_2^\pm$. 
If ${\bm d}_2^\pm=0$, 
we can use the same procedure to determine the asymptotic behavior of $Q$-factor and polarizations of resonant states near a BIC by higher order state corrections.

\section{Structures with an inversion symmetry}
In this section we show that ${\bf V}={\bf S}^{1/2}{\bf U}$ is a real matrix in a structure with an in-plane inversion symmetry.
We first construct special diffraction solutions which are linear combinations of ${\bm u}_{s*}^\pm$ and ${\bm u}_{p*}^\pm$ and satisfy
\begin{equation}\label{linearcombination}
	{\cal C}_2\left(\overline{a}_s^+\overline{\bm u}_{s*}^++\overline{a}_p^+\overline{\bm u}_{p*}^++\overline{a}_s^-\overline{\bm u}_{s*}^-+\overline{a}_p^-\overline{\bm u}_{p*}^-\right)= a_s^+{\bm u}_{s*}^++a_p^+{\bm u}_{p*}^++a_s^-{\bm u}_{s*}^-+a_p^-{\bm u}_{p*},
\end{equation}
where $a_s^\pm$ and $a_p^\pm$ are four undetermined coefficients.
Since the left and right terms in Eq.~(\ref{linearcombination}) must have the same far-field patterns, 
according to the definition of ${\bm u}_{s*}^\pm$ and ${\bm u}_{p*}^\pm$,
we have
\[
{\bf S}{\bf a}=\overline{\bf a},\quad {\bf a}=\left[
\begin{array}{c}
	a_s^+\\
	a_p^+\\
	a_s^-\\
	a_p^-
\end{array}
\right].
\]
In a structure with an inversion symmetry, 
the scattering matrix is unitary and symmetric, i.e., ${\bf S}^\dagger{\bf S}={\bf I}$ and ${\bf S}={\bf S}^{\sf{T}}$.
Denoting the eigenvalues and eigenvectors of $\bf S$ by $\lambda_j=\exp(i\phi_j)$ and ${\bf q}_j$, $j=1,2,3,4$,
we have
$\overline{\lambda}_j{\bf q}_j={\bf S}^{-1}{\bf q}_j=\overline{{\bf S}}{\bf q}_j$ 
and obtain ${\bf S}\overline{\bf q}_j={\lambda}_j\overline{\bf q}_j$.
Therefore, the eigenvectors can be scaled as real unit vectors that are still denoted by ${\bf q}_l$.
The scattering matrix $\bf S$ is diagonalizable and we have
\begin{equation}
	{\bf S}={\bf Q}\left[
	\begin{array}{cccc}
		\lambda_1& & &\\
		&\lambda_2 & &\\
		& &\lambda_3 &\\
		& & & \lambda_4
	\end{array}
	\right]{\bf Q}^{\sf{T}},\quad {\bf Q}=[{\bf q}_1,{\bf q}_2,{\bf q}_3,{\bf q}_4].
\end{equation} 
Moreover, we have
\[
{\bf S}^{\pm 1/2}={\bf Q}\left[
\begin{array}{cccc}
	\lambda_1^{\pm 1/2}& & &\\
	&\lambda_2^{\pm 1/2} & &\\
	& &\lambda_3^{\pm 1/2} &\\
	& & & \lambda_4^{\pm 1/2}
\end{array}
\right]{\bf Q}^{\sf{T}},
\]
and $\overline{\bf S}^{-1/2}={\bf S}^{1/2}$.
Let
$[{\bf a}_1, {\bf a}_2, {\bf a}_3, {\bf a}_4]={\bf S}^{-1/2}$, we obtain four diffraction solutions 
\[
{\bm u}_*^l=a_{ls}^+{\bm u}_{s*}^++a_{lp}^+{\bm u}_{p*}^++a_{ls}^-{\bm u}_{s*}^-+a_{lp}^-{\bm u}_{p*},\;l=1,2,3,4
\]
satisfying
${\cal C}_2\overline{\bm u}_*^l={\bm u}_*^l$, $1\leq l\leq 4$. 
Since $\overline{{\cal L}}_1=-{{\cal L}}_1$ and
\[
{\cal C}_2(\overline{{\cal L}_1{\bm u}_*})=-{\cal C}_2({\cal L}_1\overline{\bm u}_*)={\cal L}_1{\cal C}_2{\overline{\bm u}_*}={\cal L}_1{{\bm u}_*},
\] 
we obtain real qualities $L^2\left<{\bm u}_*^{l}|{\cal L}_{1}|{\bm u}_*\right>$.
Consequently, the matrix ${\bf V}={\bf S}^{1/2}{\bf U}=\overline{({\bf S}^{-1/2})}{\bf U}$ is real.

For Eq.~(6), if ${\bf V}$ is full rank, then ${\bf U}={\bf S}^{1/2}{\bf V}$ 
is full rank as well since the scattering matrix $\bf S$ is unitary.
For a linear system ${\bf A}{\bf x}={\bf 0}$, if ${\bf A}$ is full rank,
then the columns of $\bf A$ are linearly independent and the solution ${\bf x}$ must be zero.
Therefore, ${\bm d}_1^\pm$ cannot be zero, as $[\cos\theta\;\sin\theta]\neq 0$. 
Conversely, if $\bf V$ is rank-deficient, there exists a special angle $\theta_c$ such that the unit vector ${\bf x}=[\cos\theta_c\;\sin\theta_c]^{\sf T}$ satisfies
${\bf V}{\bf x}=0$. Since the scattering matrix $\bf S$ is unitary, then ${\bm d}_1^\pm=0$ and the ultrahigh-$Q$ resonances occur at this angle.

\section{Resonant states near an off-$\Gamma$ BIC}
\begin{figure}[htbp]
	\centering
	\includegraphics[scale=0.3]{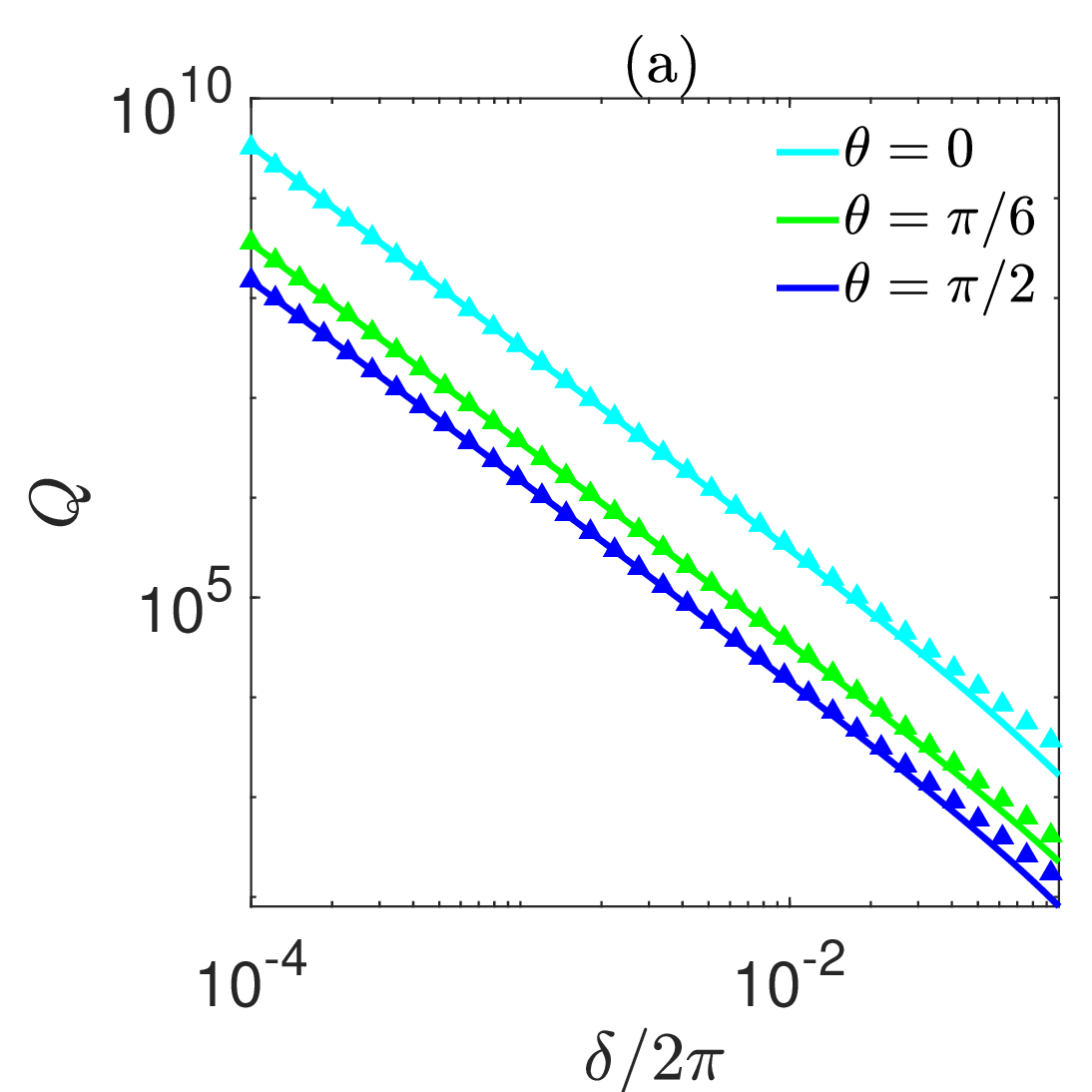}
	\includegraphics[scale=0.3]{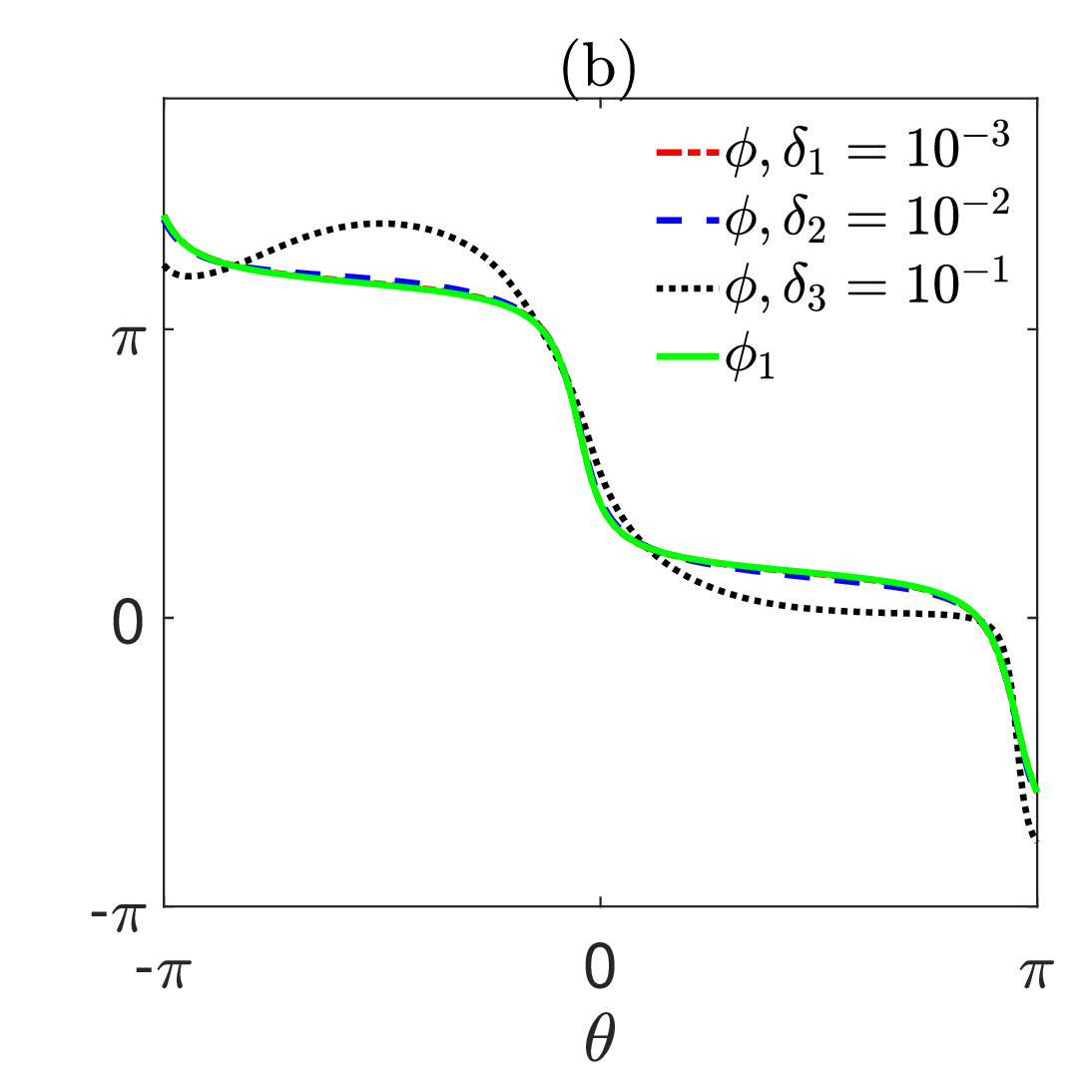}
	\includegraphics[scale=0.3]{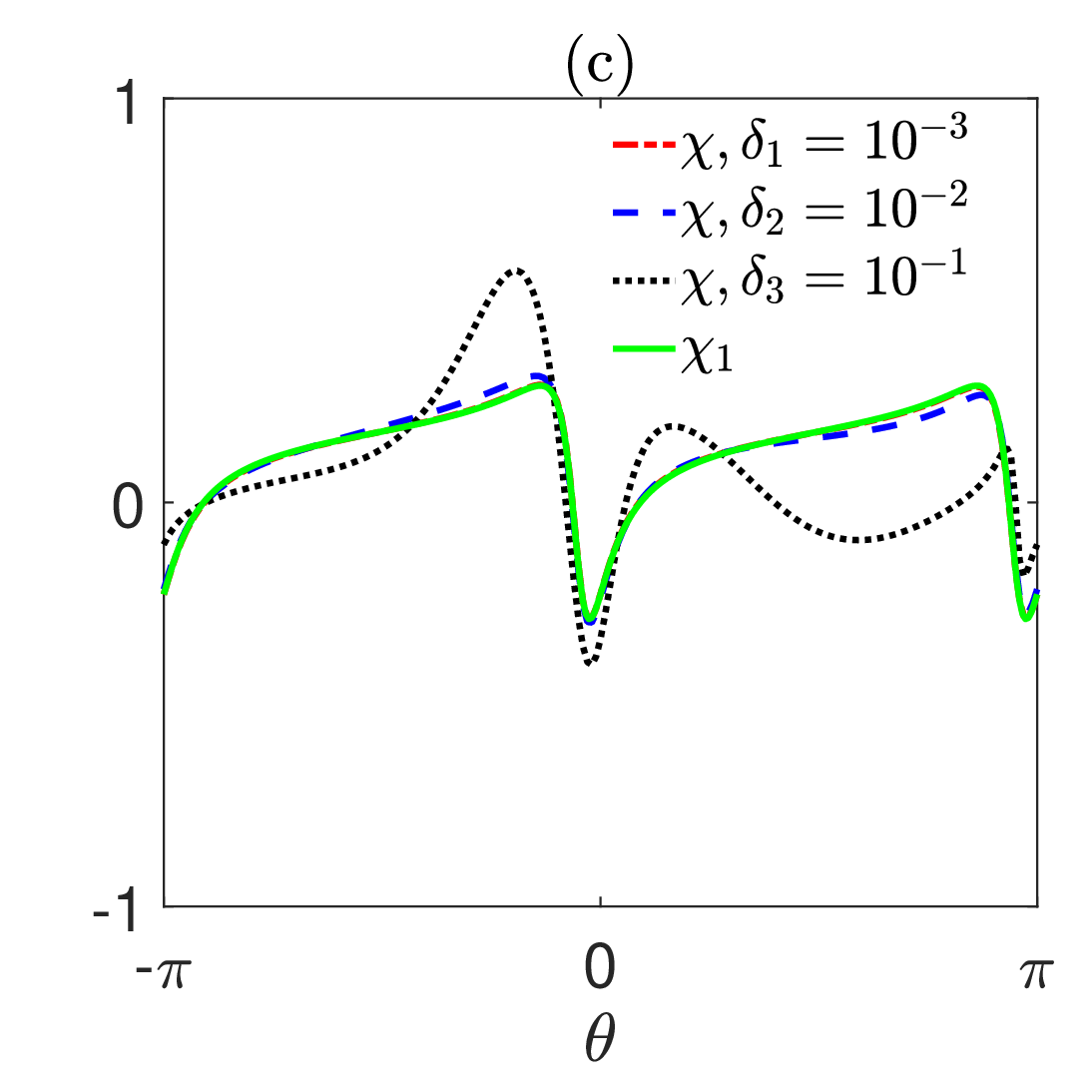}
	\caption{(a). The $Q$-factor (denoted by lines) of resonant states and the predicted results $Q_{2,\theta}$ (denoted by triangles) at three angles.
		(b) and (c). The polarization angle $\phi$ and the ellipticity $\chi$ of resonant states for three values of $\delta$. We emphasize that $Q$, $\phi$ and $\chi$ of resonant states are calculated numerically, and $Q_{2,\theta}$, $\phi_1$ and $\chi_1$ are their analytically predicted counterparts. As $\delta\rightarrow 0$, we have $\phi\sim\phi_1$ and $\chi\sim\chi_1$.
	}\label{SMPACylinderQFactor}
\end{figure}

In this section, we give an example for resonant states near an off-$\Gamma$ BIC to validate our theory.
We consider a periodic array of infinity-long circular cylinders with radius $R=0.3L$ and dielectric constant $\varepsilon_c=15$,
and study resonant states near an off-$\Gamma$ BIC with a Bloch wavevector ${\bm \alpha}_*L/2\pi=(0.1894,0.1539)$ and a free space wavenumber $kL/2\pi=0.5087$.
After calculating the scattering matrix $\bf S$ and the matrix $\bf U$, we can approximate the $Q$-factor and the far-field polarization by the polarization vector ${\bm d}_1$ calculated from Eq.~(6).
Recall that the Bloch wavevector of resonant states $\bm \alpha$ can be written as ${\bm \alpha}={\bm \alpha}_*+\delta(\cos\theta,\,\sin\theta)/L$.
We choose three angles $\theta = 0$, $\pi/4$, and $\pi/2$ and calculate $Q$-factor of resonant states with different $\delta$.
In addition, we approximate the $Q$-factor by $Q_{2,\theta}=-0.5\mbox{Re}(\omega_*)/\mbox{Im}[\delta^2\omega_2(\theta)]$.
Figure~\ref{SMPACylinderQFactor}(a) shows the $Q$-factor of resonant states (denoted by lines, calculated numerically) and the predicted results $Q_{2,\theta}$ (denoted by triangles).
It can be seen that the value of $Q_{2,\theta}$ agrees well with the $Q$-factor. 
When $\delta$ is relatively small, they are nearly identical.
Figure~\ref{SMPACylinderQFactor}(b) and (c) show the polarization angle $\phi$ and the ellipticity $\chi$ of resonant states with three values of $\delta=2\pi\delta_i$, where $i=1,2,3$ and $\delta_i=10^{i-4}$.
As $\delta\rightarrow 0$, we indeed have $\phi\sim\phi_1$ and $\chi\sim \chi_1$.
These numerical results validate our theory for the cases of off-$\Gamma$ BICs.

\section{Super-BICs in structures with $C_{4v}$ symmetry}
In this section we show that for at-$\Gamma$ super-BICs in structures with $C_{4v}$ symmetry,
we typically have $Q\sim 1/\delta^6$ in all directions.
Equivalently, we prove that ${\bm d}_1^\pm={\bm d}_2^\pm=0$ for all $\theta$.
As an illustration, we consider a non-degenerate BIC belonging to ${\it A}_2$ representation.
The four diffraction solutions ${\bm u}_{s*}^\pm$ and ${\bm u}_{p*}^\pm$ belong to $\it E$ representation~\cite{Sakoda2012OE}.
Our theory remains valid for BICs belonging to other representations.

Equation~(\ref{k1realcond}) leads to $k_1=0$ since $\left<{\bm u}_*|{\cal L}_1|{\bm u}_*\right>=0$.
According the symmetries of ${\bm u}_*$ and ${\bm u}_{l*}^\pm$, we have
$
\left<{\bm u}_{s*}^\pm|{\cal L}_{1x}|{\bm u}_*\right>=\left<{\bm u}_{p*}^\pm|{\cal L}_{1y}|{\bm u}_*\right>
$
and
$
\left<{\bm u}_{s*}^\pm|{\cal L}_{1y}|{\bm u}_*\right>=-\left<{\bm u}_{p*}^\pm|{\cal L}_{1x}|{\bm u}_*\right>.
$
Moreover, we have
\[
\left<{\bm u}_{s*}^\pm|{\cal L}_{1x}|{\bm u}_*\right>=-\left<\sigma_x{\bm u}_{s*}^\pm|\sigma_x{\cal L}_{1x}|{\bm u}_*\right>=\left<{\bm u}_{s*}^\pm|{\cal L}_{1x}\sigma_x|{\bm u}_*\right>=-\left<{\bm u}_{s*}^\pm|{\cal L}_{1x}|{\bm u}_*\right>=0,
\]
where $\sigma_x$ is a symmetry operator related to reflection symmetry in $x$.
Therefore, we have
\[
{\bf U}^\pm=L^2
\left[
\begin{array}{cc}
	0&\left<{\bm u}_{s*}^\pm|{\cal L}_{1y}|{\bm u}_*\right>\\
	\left<{\bm u}_{p*}^\pm|{\cal L}_{1x}|{\bm u}_*\right>&0
\end{array}
\right].
\]
Since cross polarization scattering cannot occur, according to the property of $\bf S$, we can show that
\[
{\bf V}=\left[
\begin{array}{cc}
	0&V_{12}\\
	-V_{12}&0\\
	0&V_{32}\\
	-V_{32}&0
\end{array}
\right].
\]

If the BIC is a super-BIC, $\bf V$ is rank-deficient and we have $V_{12}=V_{32}=0$.
Thus, ${\bm d}_1^\pm=0$ for all $\theta$.
Next, we prove that ${\bm d}_2^\pm=0$ in all directions.
The solutions ${\bm u}_1$ can be written as
\begin{equation}\label{decompositionphi1}
	{\bm u}_1={{\bm u}_{1x}}\cos\theta+{{\bm u}_{1y}}\sin\theta,
\end{equation}
where
\[
{\cal M}_*{{\bm u}_{1x}}=-{\cal L}_{1x}{\bm u}_*,\quad {\cal M}_*{{\bm u}_{1y}}=-{\cal L}_{1y}{\bm u}_*.
\]
Substituting Eq.~(\ref{decompositionphi1}) into Eq.~(\ref{LEq2}),
we have
\[
{\cal M}_*{\bm u}_2=2k_*k_2\varepsilon{\bm u}_*+({\cal L}_{1x}\cos\theta+{\cal L}_{1y}\sin\theta)(\cos\theta{{\bm u}_{1\alpha}}+\sin\theta{{\bm u}_{1\beta}})-{\cal L}_2{\bm u}_*.
\]
The vectors ${\bm d}_2^\pm$ can be calculated from Eq.~(\ref{d2sd2p}) and
\[
\begin{aligned}
	&\left<{\bm u}^\pm_{s*}|{\cal M}_*|{\bm u}_2\right>=[\cos\theta,\sin\theta]    \left[
	\begin{array}{cc}
		\left<{\bm u}_{s*}^\pm|{\cal L}_{1x}|{{\bm u}_{1x}}\right>&\left<{\bm u}_{s*}^\pm|{\cal L}_{1x}|{{\bm u}_{1y}}\right>\\
		\left<{\bm u}_{s*}^\pm|{\cal L}_{1y}|{{\bm u}_{1x}}\right>&\left<{\bm u}_{s*}^\pm|{\cal L}_{1y}|{{\bm u}_{1y}}\right>
	\end{array}
	\right]
	\left[\begin{array}{c}
		\cos\theta\\
		\sin\theta
	\end{array}
	\right]\\
	&\left<{\bm u}_{p*}^\pm|{\cal M}_*|{\bm u}_2\right>=[\cos\theta,\sin\theta]    \left[
	\begin{array}{cc}
		\left<{\bm u}_{p*}^\pm|{\cal L}_{1x}|{{\bm u}_{1x}}\right>&\left<{\bm u}_{p*}^\pm|{\cal L}_{1x}|{{\bm u}_{1y}}\right>\\
		\left<{\bm u}_{p*}^\pm|{\cal L}_{1y}|{{\bm u}_{1x}}\right>&\left<{\bm u}_{p*}^\pm|{\cal L}_{1y}|{{\bm u}_{1y}}\right>
	\end{array}
	\right]
	\left[\begin{array}{c}
		\cos\theta\\
		\sin\theta
	\end{array}
	\right].
\end{aligned}
\]
Since ${{{\cal C}_2}}{\cal L}_{1\sigma}{{\bm u}_{1\kappa }}={\cal L}_{1\sigma}{{\bm u}_{1\kappa}}$, $\sigma,\kappa\in\{x,y\}$,
the matrices shown in the above formulas are zero.
Thus, ${\bm d}_2^\pm=0$ and $Q\sim 1/\delta^6$ for all $\theta$ at least.
For super-BICs in a structure with an inversion symmetry (e.g., a periodic array of cylinders), our theory applies to this case as well.

\begin{figure}[htbp]
	\centering
	\includegraphics[scale=0.38]{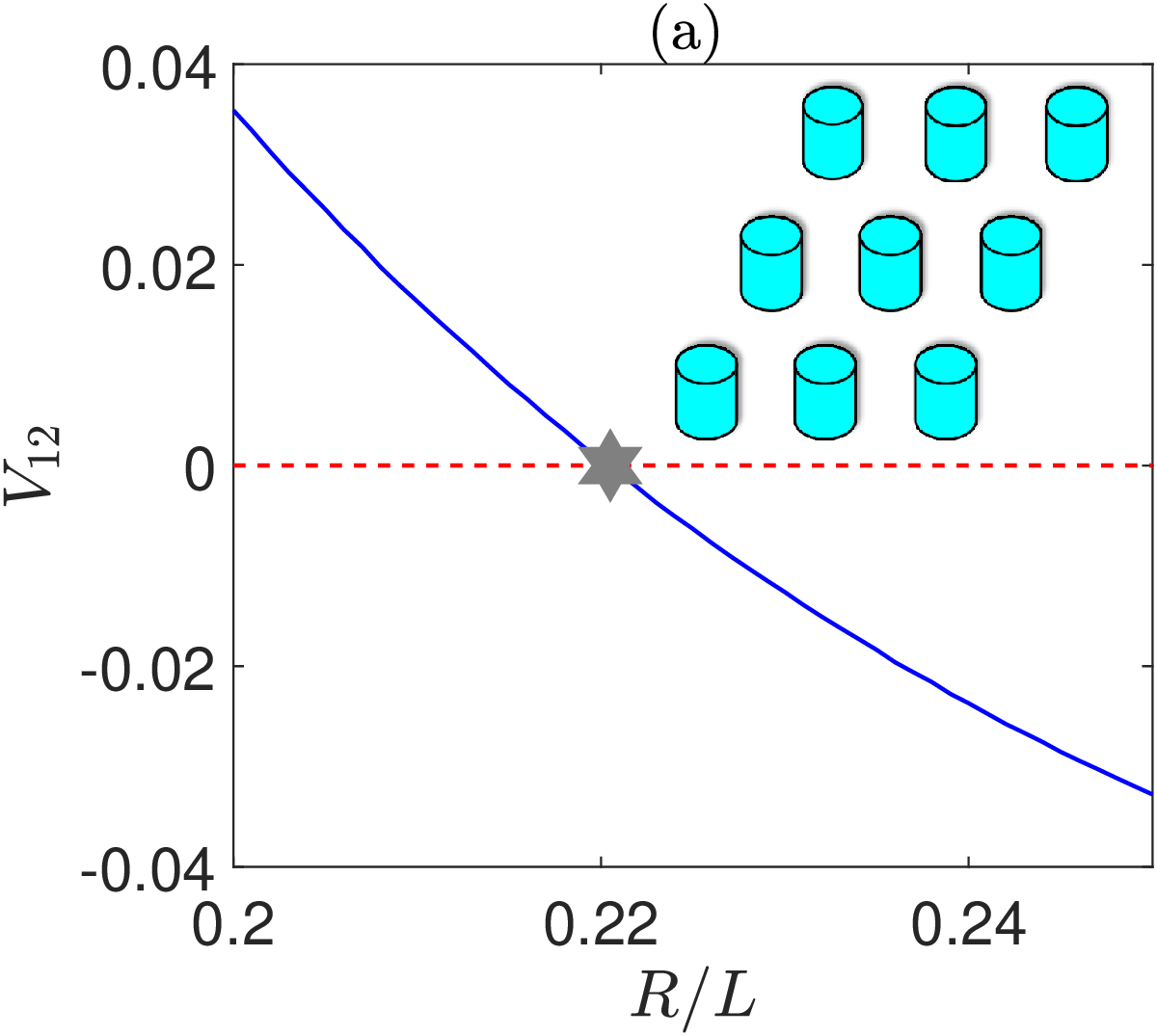}\quad
	\includegraphics[scale=0.38]{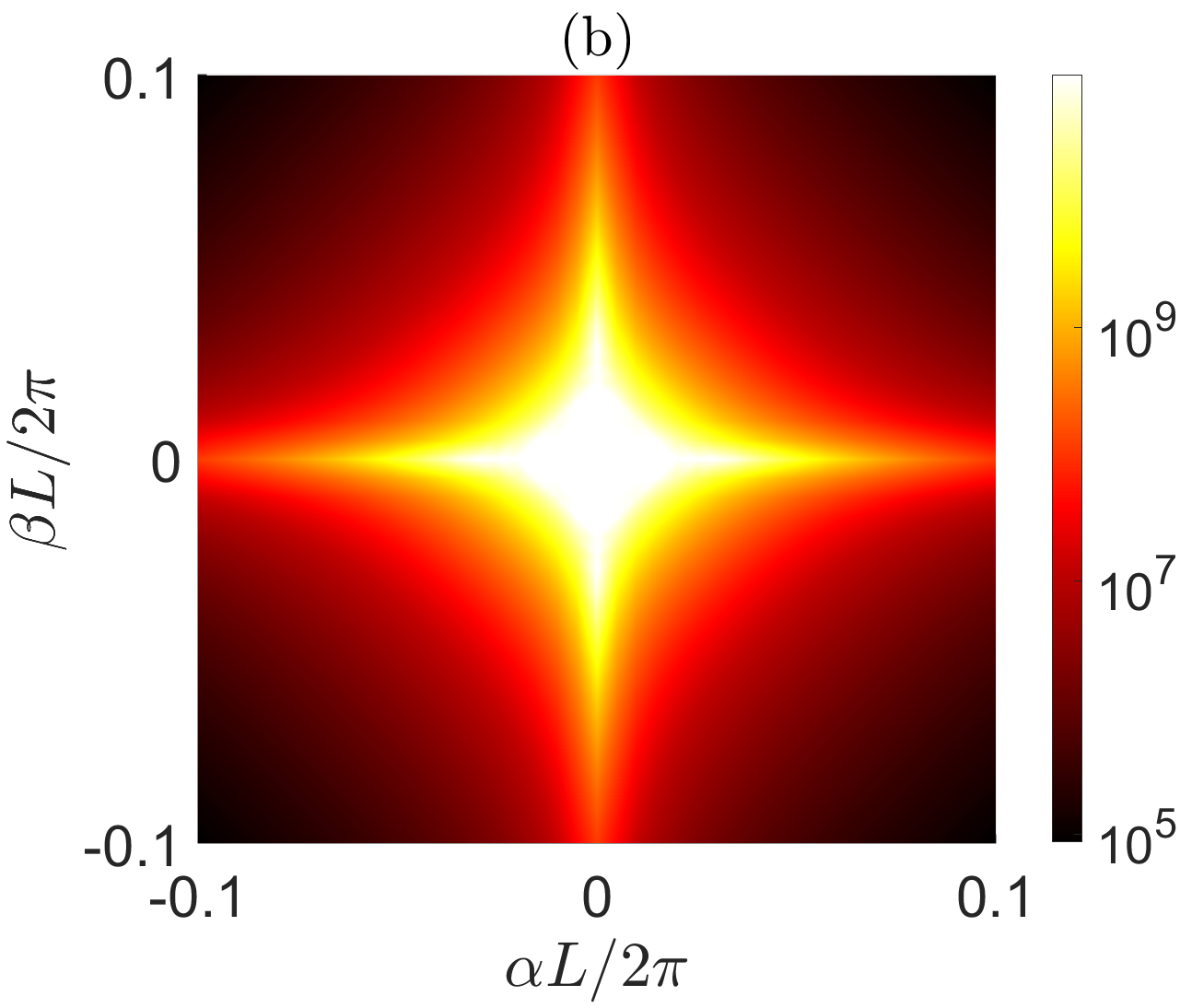}\\
	\vspace{0.15cm}
	\includegraphics[scale=0.38]{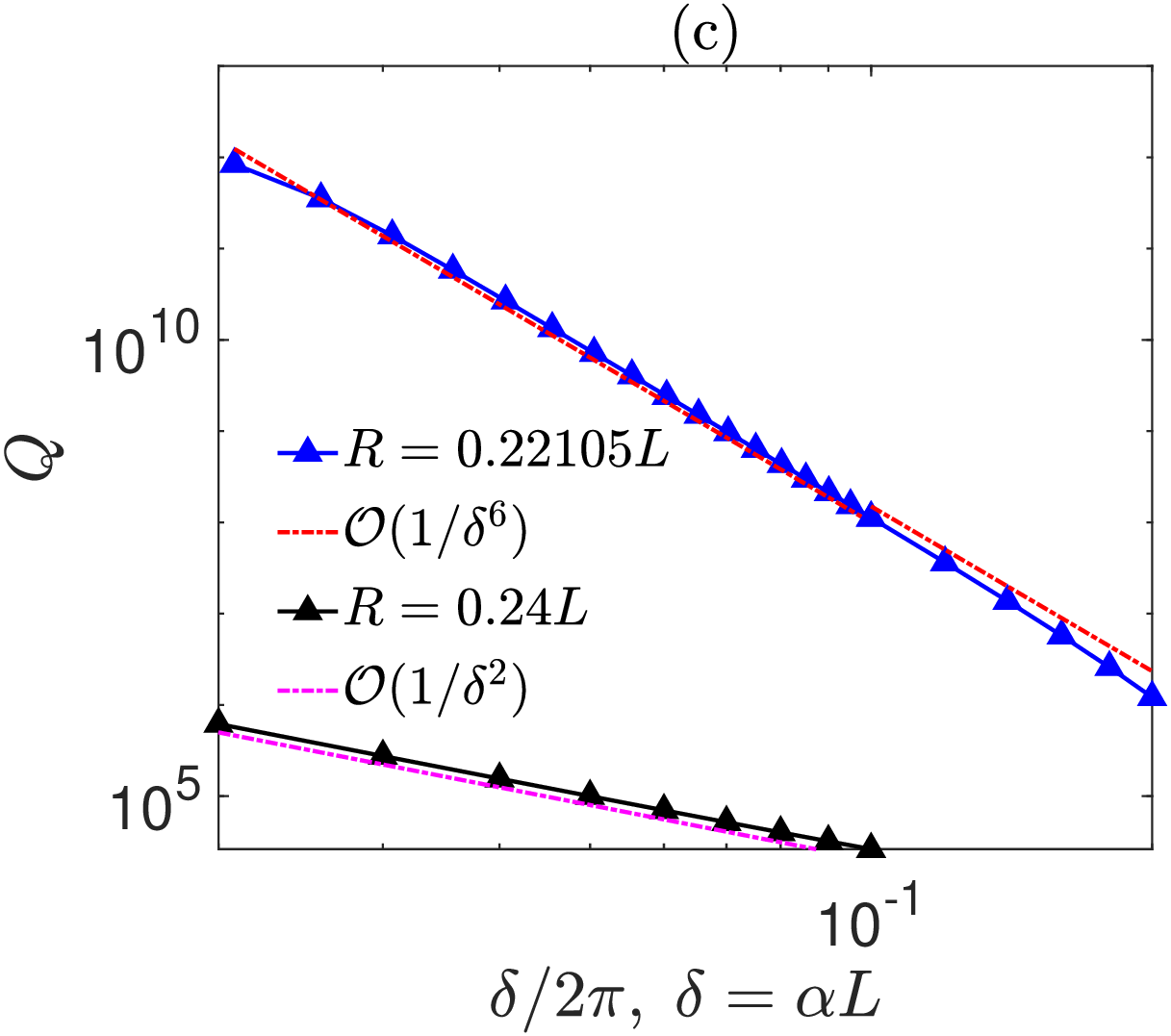}\quad
	\includegraphics[scale=0.38]{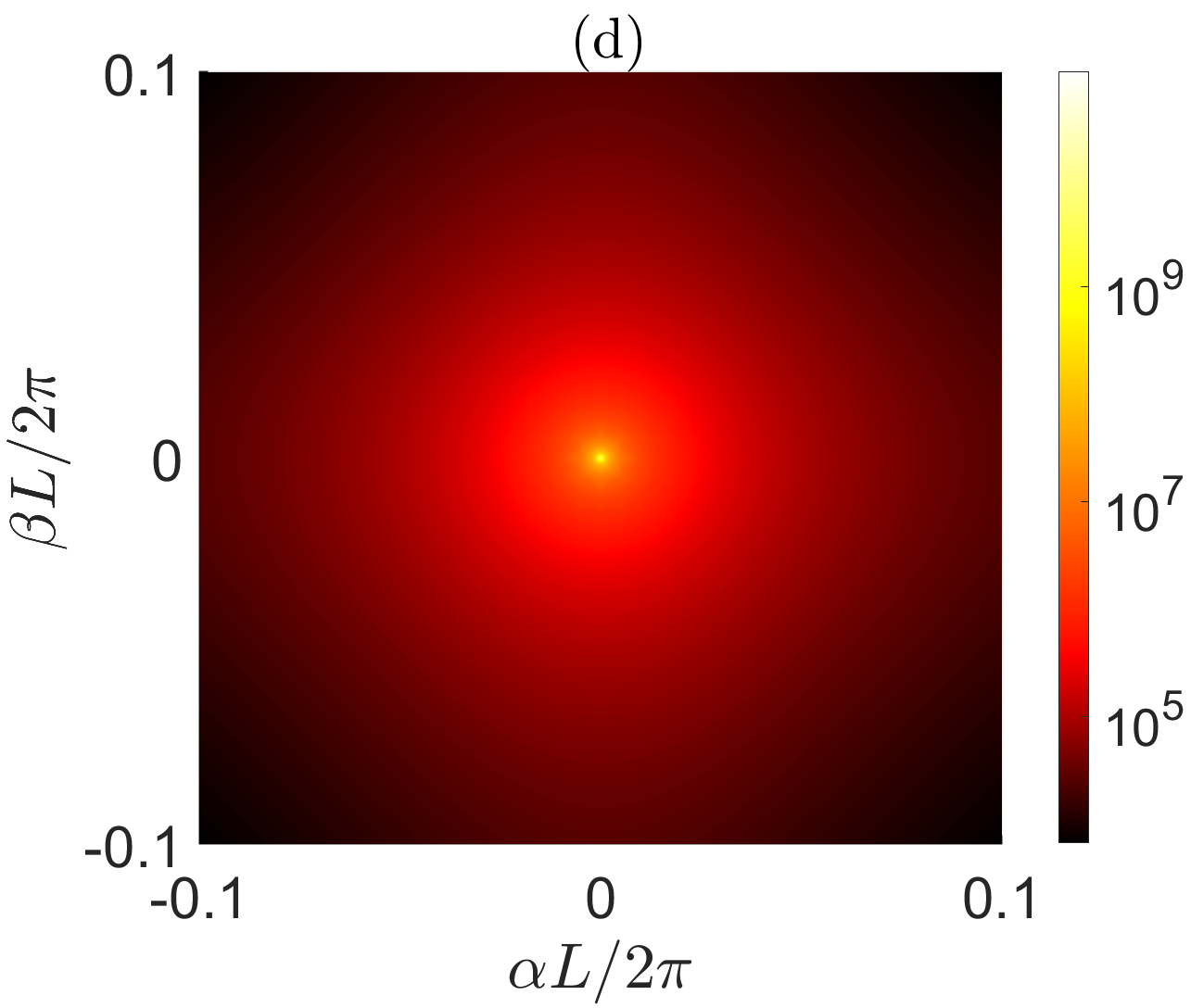}
	\caption{Super-BICs in a square lattice of silicon rods in air.
		(a). The quantity $V_{12}$ as a function of radius $R$.
		(b). $Q$-factor (calculated numerically) of resonant states near a super-BIC ($\omega_*L/2\pi c=0.541$) at $R=0.221L$.
		(c). Asymptotic behavior of $Q$-factor.
		(d). $Q$-factor (calculated numerically) of resonant states near a generic BIC at $R=0.24L$.}\label{SMSymSuperBIC}
\end{figure}

To validate our theory, we consider a square lattice of finite-length silicon circular cylinders ($\varepsilon_{\rm Si}=12$) embedded in air.
We have $V_{12}=V_{32}$ due to up-down mirror symmetry and only need to tune one parameter to obtain a zero of $V_{12}$.
The height $h=0.8L$ is fixed and the quantity $V_{12}$ as a function of radius $R$ is shown in Fig.~\ref{SMSymSuperBIC}(a).
It can be seen that at $R=0.221L$, we have $V_{12}=0$ and obtain a super-BIC (marked by a gray hexagram). 
As indicated in Fig.~\ref{SMSymSuperBIC}(b)-(c), we indeed have $Q\sim 1/\delta^6$.
We also show the $Q$-factor of resonant states near a generic BIC at $R=0.24L$ in Fig.~\ref{SMSymSuperBIC}(c) and (d).
It is obvious that super-BICs can improve the $Q$-factor in a broad wavevector range.
\section{Nearly linear and circular polarizations of resonant states and topological natures of BICs}
In this section, we study polarizations of resonant states near a generic BIC in structures with an in-plane inversion and up-down mirror symmetries.
We determine BICs and related angles where nearly linear or circular polarizations can occur.
If near-circular polarization cannot occur for all angles, 
we show that the topological charge must be $\pm 1$.

The polarizations of resonant states are nearly linear at some angles if $\mbox{Im}(\overline{d}_{1s}{d}_{1p})=0$,
where $d_{1s}$ and $d_{1p}$ are calculated from Eq.~(6) with the scattering matrix $\bf S$ given by
\[
{\bf S}=\left[
\begin{array}{cc}
	C_{ss}&C_{ps}\\
	C_{sp}&C_{pp}
\end{array}
\right].
\]
This condition is equivalent to 
\[
{\bf v}^{\sf T}\mbox{Im}\left(\left[
\begin{array}{c}
	\overline{S}_{11}\\
	\overline{S}_{12}		
\end{array}
\right]\left[
\begin{array}{cc}
	{S}_{21}&S_{22}
\end{array}
\right]\right){\bf v}=0,
\]
where
\[
{\bf v}={\bf V}\left[\begin{array}{c}
	\cos\theta\\
	\sin\theta
\end{array}
\right],\;
{\bf S}^{1/2}=\left[
\begin{array}{cc}
	{S}_{11}&S_{12}\\
	{S}_{21}&S_{22}		
\end{array}
\right].
\]
Since the scattering matrix depends solely on BICs, the verification for this condition is straightforward.
The matrix $\bf S$ is unitary and symmetric, and can be written as
\[
{\bf S}=e^{i{\kappa}}\left[
\begin{array}{cc}
	{\varrho} e^{i{\psi}}&\sqrt{1-{\varrho}^2}\\
	\sqrt{1-{\varrho}^2}&-{\varrho} e^{-i{\psi}}
\end{array}
\right],
\]
where $0\leq {\varrho}\leq 1$ and $0\leq {\kappa},{\psi}< 2\pi$.
Moreover, the matrix ${\bf S}^{1/2}$ is also unitary and symmetric, and has the form of
\[
{\bf S}^{1/2}=e^{i\tilde{\kappa}}\left[
\begin{array}{cc}
	\tilde{\varrho} e^{i\tilde{\psi}}&\sqrt{1-\tilde{\varrho}^2}\\
	\sqrt{1-\tilde{\varrho}^2}&-\tilde{\varrho} e^{-i\tilde{\psi}}
\end{array}
\right],
\]
where $0\leq \tilde{\varrho}\leq 1$ and $0\leq \tilde{\kappa},\tilde{\psi}< 2\pi$. 
Therefore,
it can be demonstrated that if $\tilde{\psi}=0$ or $\pi$, all the polarization states near the BIC are nearly linear.
This condition is equivalent to $C_{ss}={C}_{pp}$ and $C_{sp}=C_{ps}=0$.
We can show that at-$\Gamma$ generic BICs in structures with $C_{4v}$ symmetry satisfy this condition. 

Next we give the condition for the existence of near-circular polarizations. Since
\[
d_{1s}^2+d_{1p}^2={\bf v}^{\sf T}{\bf S}{\bf v},
\]
we show that $d_{1s}^2+d_{1p}^2=0$ at an angle $\theta_c$ if and only if
\begin{equation}\label{cond1}
	{\bf v}_c^{\sf{T}}\left[
	\begin{array}{cc}
		\varrho \cos\psi&\sqrt{1-\varrho^2}\\
		\sqrt{1-\varrho^2}&-\varrho \cos\psi
	\end{array}
	\right]	{\bf v}_c=0,\quad
	\varrho\sin\psi	\|{\bf v}_c\|_2^2=0,
\end{equation}
where
\[
{\bf v}_c={\bf V}\left[
\begin{array}{c}
	\cos\theta_c\\
	\sin\theta_c
\end{array}
\right].
\]
On the other hand, Eq.~(\ref{cond1}) have nonzero solutions ${\bf v}_c$ if and only if $\varrho\sin\psi=0$.
Therefore, for a generic BIC, there exist special angles at which ${\bm d}_1$ is circularly polarized if and only if $\varrho\sin\psi=0$,
i.e., $C_{ss}=C_{pp}=0$ or $\arg(C_{ss}\overline{C}_{pp})=\pm \pi$.
These angles can be solved from Eq.~(\ref{cond1}).

Next, we show that if near-circular polarizations cannot occur for all angles,
the topological charge of the BIC must be $\pm 1$.
The matrix $\bf F$ and the vector $\bf g$ given in Eq.~(10) can be obtained from Eq.~(6) and the definition of Stokes parameters directly.
A straightforward calculation shows that the determinant of the matrix ${\bf F}$ is
\[
\det {\bf F}=-0.5\det{\bf V}(1-2\tilde{\varrho}^2\sin^2\tilde{\psi})\|{\bf V}\|_F^2,
\]
where $\|{\bf V}\|_F$ is Frobenius norm of the matrix $\bf V$.
For a generic BIC, both $\det {\bf V}$ and $\|{\bf V}\|_F$ are nonzero.
In addition, it can be shown that $(1-2\tilde{\varrho}^2\sin^2\tilde{\psi})=0$ if and only if $\varrho\sin\psi=0$ since ${\bf S}^{1/2}{\bf S}^{1/2}={\bf S}$. 
Therefore, if near-circular polarization cannot occur, we must have $\det {\bf F}\neq 0$.  
In this case, we can further show that $\|{\bf F}^{-1}{\bf g} \|_2<1$. 
Consequently, the trajectory $(\mathbb{S}_{1,1},\mathbb{S}_{1,2})$, $\theta\in[-\pi,\pi)$ forms an ellipse with two revolutions around the origin. 
The revolution direction depends on the sign of $\det {\bf F}$ (or $\bf S$ and $\bf V$).

\section{Numerical approach for calculating resonant states and super-BICs}
In this Letter, we utilize the Dirichlet-to-Neumann method introduced in Ref.~\cite{LJYuan2017} 
and the transverse impedance operator method developed in Ref.~\cite{Zhang2024arXiv} to calculate resonant states in
a periodic array of infinity-long circular cylinders and PhC slabs, respectively.
A BIC is a specific type of resonant state characterized by a real frequency.
Both methods result in a nonlinear eigenvalue problem expressed as
\[
{\bf T}(\omega){\bf h}=0,
\]
where $(\omega,{\bf h})$ represents an eigenpair corresponding to the frequency and certain quantities of the electromagnetic fields. 
The scattering states associated with BICs can be determined from the linear systems
\[
{\bf T}(\omega_*){\bf h}_s={\bf f}_{\rm inc},
\]
where $\omega_*$ denotes the frequency of the BIC, and the right hand side ${\bf f}_{\rm inc}$ is associated with the incident wave.

In structures with in-plane inversion symmetry, a super-BIC has a rank-deficient matrix $\bf V$ with the singular values $\sigma_1({\bf V})\geq\sigma_{2}({\bf V})=0$.
Consequently, identifying such a super-BIC is equivalent to finding the zeros of the vector function ${\bm f}(\omega, {\bm\alpha}, {\bm p})$ given by
\[
{\bm f}=\left[
\begin{array}{c}
	\lambda_{\rm \min}({\bf T})\\
	\sigma_1\sigma_2
\end{array}
\right],
\] 
where ${\bm p}$  represents a set of structural parameters.
If iterative methods, such as Newton method, are used to find the zeros, the function $\bm f$ must be computed for given values of $\omega$, $\bm \alpha$ and $\bm p$. 
In the following, we consider a metasurface illustrated in Fig. 2(a) and apply the method presented in Ref.~\cite{Zhang2024arXiv} to calculate the function ${\bm f}$.

Let $D>h/2$ and define the planes $\Gamma_\pm=S\times \{z=\pm D\}$, where $S=(-L/2,L/2)\times(-L/2,L/2)$.
We divide the unit cell $\Omega$ into three subdomains $\Omega_+$, $\Omega_-$ and $\Omega_D$:
\[
\Omega_+=S\times(D,+\infty),\quad\Omega_-=S\times(-\infty,-D),\quad \Omega_D=S\times(-D,D).
\]
Since $\varepsilon^\pm$ are constants in the regions $\Omega_\pm$,
the periodic state ${\bm u}_*$ of a BIC can be expressed using Rayleigh expansion as follows:
\begin{equation}\label{EXP}
	{\bm u}_*\approx\sum_{m,n=-M}^{M}{\bm u}_{*,mn}^{\pm}e^{i2\pi(mx+ny)/L\pm i\gamma^\pm_{mn}z},\;\pm z>D,
\end{equation} 
where ${\bm u}_{*,mn}^{\pm}$ are Fourier coefficients. 
The coefficients ${\bm u}_{*,00}^\pm$ are zero since the corresponding term can propagate and BICs do not radiate power.
In the fields $\Omega_D$, the states ${\bm u}_*$ is expanded by finite element basis functions
\begin{equation}\label{FEME}
	{\bm u}_*\approx\sum_{j=1}^{N}c_{*j}{\bm N}_j,
\end{equation}
where $N$ is the number of the basis functions $\{{\bm N}_j\}_{j=1}^N$.
The coefficients ${\bm u}_{*,mn}^{\pm}$ and $c_{*j}$ can be derived  from the vector ${\bf h}$~\cite{Zhang2024arXiv}.
The scattering states have analogous expansions, and we provide the form for $\tilde{\bm u}_{s*}^+$:
\begin{equation}\label{EXPS}
	\tilde{\bm u}_{s*}^+\approx\left\{
	\begin{aligned}
		&\displaystyle \eta^+\hat{s}_*e^{-i\gamma^+_{mn}z}+\sum_{m,n=-M}^{M}{\bm u}_{s*,mn}^{+}e^{i2\pi(mx+ny)/L+ i\gamma^+_{mn}z},\;\text{ in }\Omega_+\\
		&\displaystyle \sum_{m,n=-M}^{M}{\bm u}_{s*,mn}^{-}e^{i2\pi(mx+ny)/L- i\gamma^-_{mn}z},\;\text{ in }\Omega_-\\
		&\sum_{j=1}^{N}c_{sj}^+{\bm N}_j,\;\text{ in }\Omega_D.
	\end{aligned}
	\right.
\end{equation}
The coefficients ${\bm u}_{s*,mn}^{\pm}$ and $c_{sj}^+$ can also be obtained from the vector ${\bf h}_{s}$~\cite{Zhang2024arXiv}.
The steps for calculating the vector function ${\bm f}$ are outlined  as follows:

Step 1: For given $\omega$, ${\bm \alpha}$, and $\bm s$, compute the matrix $\bf T$ using the method developed in Ref.~\cite{Zhang2024arXiv}.

Step 2: Calculate the minimum module eigenvalue $\lambda_{\rm min}({\bf T})$ and eigenvector $\bf h$.
Normalize and scale the field such that $\left<{\bm u}_*|\varepsilon|{\bm u}_*\right>=1$ and ${\cal C}_2\overline{\bm u}_*={\bm u}_*$.

Step 3: Calculate the scattering states ${\bm u}_{l*}^\pm$ and determine  the scattering matrix $\bf S$, where $l\in\{s,p\}$.
Adjust the scattering states to ensure orthogonality with the BIC, i.e., $\left<{\bm u}_{l*}^\pm|\varepsilon|{\bm u}_*\right>=0$.

Step 4: Compute  the matrix $\bf U$ (elements $\left<{\bm u}_{l*}^\pm|{\cal L}_{1\sigma}|{\bm u}_*\right>$, $\sigma\in\{x,y\}$) and then obtain ${\bf V}={\bf S}^{1/2}{\bf U}$,
along with $\sigma_1({\bf V})$ and $\sigma_2({\bf V})$.

The integral $\left<{\bm w}|{\cal A}|{\bm u}_*\right>$ arising from steps 2, 3 and 4
can be decomposed into
\[
\left<{\bm w}|{\cal A}|{\bm u}_*\right>=C_++C_-+C_D,
\]
where ${\bm w}={\bm u}_*$ or ${\bm u}_{l*}^\pm$, ${\cal A}={\cal L}_1$ or $\varepsilon$, and
\[
C_\pm = \int_{\Omega_\pm}\overline{\bm w}\cdot{\cal A}{\bm u}_*\,{\rm d}x{\rm d}y{\rm d}z,\; C_D = \int_{\Omega_D}\overline{\bm w}\cdot{\cal A}{\bm u}_*\,{\rm d}x{\rm d}y{\rm d}z.
\]
These quantities can be directly computed using Eqs.~(\ref{EXP}), (\ref{FEME}) and (\ref{EXPS}).
Super-BICs presented in Figs. 1 and 2 in the Letter are calculated by the aforementioned algorithm. 
The set of structural parameters $\bm p$ includes the height $h$ and the radius $R$ of the dielectric rods.
For these examples, since $\bf V$ assumes a simple form, it is unnecessary to compute its singular values. 
For the super-BICs illustrated in Figs. 2 and S2, the function $\bm f$ can be expressed as
\[
{\bm f}=\left[
\begin{array}{c}
	\lambda_{\rm \min}({\bf T})\\
	V_{12}\\
	V_{32}
\end{array}
\right].
\] 
We employ Newton's method to determine the zeros of $\bm f$.
Alternatively, we can first compute BICs for varying heights and radii, subsequently calculating 
$V_{12}$ and $V_{32}$ to identify the zeros of $\bm f$. 
Both methods yield the same super-BICs, although the former is more efficient than the latter.

\end{document}